\documentclass{aa}  

\usepackage{graphicx}
\usepackage{nicefrac}
\usepackage{xcolor}
\usepackage{lipsum}
\usepackage{caption}
\usepackage{subcaption}
\usepackage{dblfloatfix}    
\usepackage{txfonts}
\usepackage[hidelinks]{hyperref}

\newcommand{\aperp}{{\alpha_\perp}}	
\newcommand{\apara}{{\alpha_\parallel}}

\newcommand{\hmpc }{$h^{-1}$Mpc}

\begin{document}

   \title{Baryon acoustic oscillations from a joint analysis of the large-scale clustering in Fourier and configuration space}

   \author{Tyann Dumerchat
          \inst{\ref{amu}}
          \and
          Julian E. Bautista
          \inst{\ref{amu}}
          }

   \institute{Aix Marseille Univ, CNRS/IN2P3, CPPM, Marseille, France \\
              \email{dumerchat@cppm.in2p3.fr, bautista@cppm.in2p3.fr}
              \label{amu}
             }


 
  \abstract
    {Baryon acoustic oscillations (BAOs) are a powerful probe of the expansion 
    history of our Universe and are typically measured in the two-point 
    statistics of a galaxy survey, either in Fourier space or in configuration space. 
    In this work, we report a first measurement of BAOs from a joint fit of 
    power spectrum and correlation function multipoles.  
    We tested our new framework with a set of 1000 mock catalogs and 
    showed that our method yields smaller biases on BAO parameters than 
    individually fitting power spectra or correlation functions, or 
    when combining them with the Gaussian approximation method. 
    Our estimated uncertainties are slightly larger than 
    those from the Gaussian approximation, likely due to noise in our sample covariance matrix, 
    the larger number of nuisance parameters, or the fact that our new framework 
    does not rely on the assumption of Gaussian likelihoods for the BAO parameters.
    However, we argue that our uncertainties are more reliable since they rely 
    on fewer assumptions, and because our method takes correlations 
    between Fourier and configuration space at the level of the two-point statistics. 
    We performed a joint analysis of the luminous red galaxy sample
    of the extended baryon oscillation spectroscopic survey (eBOSS) data release 16, 
    obtaining $D_H/r_d = 19.27 \pm 0.48$ and $D_M/r_d = 17.77 \pm 0.37$, in excellent
    agreement with the official eBOSS consensus BAO-only results 
    $D_H/r_d = 19.33 \pm 0.53$ and $D_M/r_d =17.86 \pm 0.33$.}

   \keywords{cosmology: cosmological parameters – cosmology: large-scale structure of the
Universe}

    \titlerunning{BAO from a joint Fourier and configuration-space analysis}
    \authorrunning{T. Dumerchat \& J. Bautista}
    
   \maketitle
%

\section{Introduction}

The study of the accelerated nature of the expansion of the Universe has
seen significant progress in the past decade thanks to measurements of
baryon acoustic oscillations (BAOs) in the three-dimensional 
distribution of galaxies. Similarly to type-Ia supernovae (see 
\citealt{broutPantheonAnalysisCosmological2022} and references therein 
for the latest results), 
BAO measurements have shown that 
the expansion is accelerating, imposing the need of a dark energy component
in the energy budget of the Universe. The next decade will see a large flow
of data coming from galaxy surveys whose goal is the precise measurement 
of BAO over a large span of the cosmic history. 

The latest BAO measurements from spectroscopic surveys were produced by 
the cosmological component of the fourth generation of the Sloan Digital 
Sky Survey (SDSS-IV, \citealt{blantonSloanDigitalSky2017}), named 
the extended Baryon Oscillation Spectroscopic Survey (eBOSS, 
\citealt{dawsonSDSSIVExtendedBaryon2016}). The eBOSS project produced measurements
of BAO using luminous red galaxies at an effective redshift $z_{\rm eff} = 0.7$
\citep{BautistaCompletedSDSSIVExtended2021, GilMarinCompletedSDSSIVExtended2021},
using emission line galaxies at $z_{\rm eff} = 0.85$ \citep{raichoorCompletedSDSSIVExtended2020, tamoneCompletedSDSSIVExtended2020, demattiaCompletedSDSSIVExtended2021}, using quasars as tracers of the matter field 
at $z_{\rm eff} = 1.48$ \citep{houCompletedSDSSIVExtended2021, neveuxCompletedSDSSIVExtended2020}, and quasars with visible Lyman-$\alpha$
forests at $z_{\rm eff} = 2.33$ \citep{dumasdesbourbouxhelionCompletedSDSSIVExtended2020}. 
Two additional 
lower redshift measurements were made at $z_{\rm eff} = 0.38$ and 0.51 
from the third generation of SDSS, BOSS \citep{eisensteinSDSSIIIMassiveSpectroscopic2011,dawsonBaryonOscillationSpectroscopic2013, alamClusteringGalaxiesCompleted2017}. At $z_{\rm eff} = 0.15$, we
have measurements from the SDSS-II Main Galaxy Sample 
\citep{rossClusteringSDSSDR72015, howlettClusteringSDSSMain2015}. 
Measurements from other surveys include the 
6dFGS \citep{beutler_6df_2011} at $z_{\rm eff} = 0.10$ and  
the WiggleZ Dark Energy Survey \citep{kazin_wigglez_2014} with three redshift bins 
spanning $0.2 < z < 1.0$.
The latest BAO measurements from photometric surveys were produced by
the Dark Energy Survey (DES, 
\citealt{darkenergysurveycollaborationDarkEnergySurvey2016}). 
DES produced angular BAO 
measurements in five redshift intervals between 0.6 and 1.1 using three years
of its data \citep{descollaborationDarkEnergySurvey2021a}.

Baryon acoustic oscillation measurements have traditionally been performed in 
configuration space and/or in Fourier space. 
Configuration space analysis is based on estimates of correlation function 
$\xi$ as a function of separation $\vec{r}$, 
commonly done with pair-counting techniques. 
Fourier space analyses assign galaxies into a regular grid
so as to Fourier transform it in order to compute the power spectrum multipoles $P_\ell$ as
a function of wavevector $\vec{k}$.
Both types of analyses perform a statistical measurement
starting from the exact same dataset, defined as a list of angular 
positions, redshifts, and weights, as well as some definition of the 
window function of the survey which is often a purely Poisson set of points 
following the same angular and redshift distribution as the real data. 
In principle, both types of analyses should yield the same cosmological
constraints since they have a common starting point. In practice,
choosing a limited range of scales used when fitting models slightly breaks
this perfect degeneracy. Noise properties or systematic effects on the 
estimated (and binned) statistics also differ between 
$\xi(\vec{r})$ and $P(\vec{k})$. 
The models of clustering used to fit two-point functions also commonly differ,
that is to say when performing a full shape analysis the clustering models in configuration space can differ from just a Fourier transform of
those in Fourier space. All these differences slightly reduce the 
correlations to levels below 100 per cent. As an example, 
\citet{BautistaCompletedSDSSIVExtended2021} show using the best-fit values for the dilation parameters on an ensemble of mock catalogues 
that BAO results in configuration and Fourier spaces are roughly 90 per cent 
correlated, for their particular choices of scales and models.
Recent analysis of BAO and redshift-space distortions (RSD) usually combine
results from configuration and Fourier space analyses into a single consensus
result. 

Given that the correlations between cosmological results 
are not exactly 100 per cent between configuration and Fourier space analyses, 
there is a slight statistical gain in producing a combined result.
Systematic errors might as well be reduced by this combination, 
since they may manifest themselves differently in both spaces.
In collaboration working groups, often two or more teams produce their analyses in a single space, 
making it difficult to decide which results to quote as final. 
Therefore, there are several advantages to produce joint results between
Fourier and configuration spaces.

When calculating a consensus result, it is important to correctly take into
account the strong correlations between measurements in configuration and 
Fourier space. \citet{sanchezClusteringGalaxiesCompleted2017a} described a 
method to compute a consensus result, which was used for the first time 
in SDSS-III BOSS \citep{alamClusteringGalaxiesCompleted2017} and extended to 
the latest measurements from SDSS-IV eBOSS survey. 
This method was used to produce consensus results on BAO and RSD analyses
of a given survey, yielding constraints on $D_H/r_d$, $D_M/r_d$ and $f\sigma_8$,
where 
$D_H$ is the Hubble distance $c/H$, 
$D_M$ is the comoving angular diameter distance, 
$r_d$ is the comoving sound horizon at drag epoch (the BAO scale), 
$f$ is the growth-rate of structures and 
$\sigma_8$ is the normalization of the smoothed linear matter power spectrum.
However, the method 
from \citet{sanchezClusteringGalaxiesCompleted2017a} assumes that the 
individual likelihoods on BAO parameters are Gaussian, which is not 
necessarily true, particularly in a regime of low signal-to-noise ratio.

In this work, we developed a framework to perform simultaneous BAO analyses 
in both configuration and Fourier spaces. 
The advantages of our framework over past work are
1) individual likelihoods on BAO parameters from each space are not assumed to be Gaussian,
2) the resulting posterior distribution is not necessarily Gaussian, resulting
in more reliable uncertainties,
3) it is simpler, and
4) yields smaller systematic biases. 
Our method only relies on a sufficiently good estimate of the full 
covariance matrix, particularly the cross-covariance between two-point functions
in Fourier and configuration spaces. We validated our method 
on realistic mock catalogues and performed comparison with previous work. 

This paper is organized as follows. 
In section~\ref{sec:data}, we describe the dataset used to validate our methods. 
In section~\ref{sec:method}, we introduce the BAO modeling, which is mostly the
same used in \citet{BautistaCompletedSDSSIVExtended2021, GilMarinCompletedSDSSIVExtended2021}, as well as the methods to produce consensus 
results.
Section~\ref{sec:results_mocks} presents several statistical and systematical tests
of both methods and finally the application on real 
data on section~\ref{sec:results_data}.

\section{Dataset}
\label{sec:data} 

To validate our methodology, we used 1000 mock catalogues reproducing 
the Luminous Red Galaxy (LRG) sample from the eBOSS survey, though 
our methods are not survey specific. This dataset is the same employed 
in the cosmological analyses of \citet{BautistaCompletedSDSSIVExtended2021, GilMarinCompletedSDSSIVExtended2021}, and we refer the reader to these
references for further details. 

\subsection{The eBOSS survey}
\label{sec:data:eboss}

The extended Baryon Oscillation Spectroscopic Survey (eBOSS,
\citealt{dawsonSDSSIVExtendedBaryon2016}) was a 5-year 
observing program using multiobject fiber-fed spectrographs
\citep{smeeMultiobjectFiberfedSpectrographs2013} 
mounted on the focal plane of the 2.5 meter Sloan Foundation 
Telescope \citep{gunnTelescopeSloanDigital2006} at the Apache Point Observatory.
During eBOSS, 174,816 LRG redshifts were obtained over 4,242~deg$^{2}$ 
of both northern and southern skies, in the redshift interval $0.6<z<1.0$. 
These were combined with BOSS galaxies in the same redshift range covering 
9,493~deg$^{2}$ of sky, resulting in a total of 377,458 LRG redshifts
\citep{rossCompletedSDSSIVExtended2020}. 
The survey geometry was defined using a set of randomly distributed points,
taking into account masked areas. The final data sample has correction 
weights accounting for the photometric and spectroscopic incompleteness, 
as well as some spurious correlations (e.g., with stellar density, Galactic
extinction). 
The final catalogue is the basis for mock catalogue production.

\subsection{Mock catalogues}
\label{sec:data:mocks}

In this work we use a sample of 1000 mock realizations of the eBOSS LRG sample
to validate our methodology and to perform statistical tests. 
A set of 1000 realisations of the eBOSS LRG survey were produced using the 
\textsc{EZmock} method \citep{chuangEZmocksExtendingZel2015}, which employs the 
Zel'dovich approximation to compute the density field at a given redshift. 
This method is faster than n-body simulations and has been calibrated to 
reproduce the two- and three-point statistics of the given galaxy sample, 
to a good approximation and up to mildly nonlinear scales. 
The angular and redshift distributions of the eBOSS LRG sample were reproduced, 
including both photo and spectro-incompleteness effects as well as systematic 
effects introducing spurious correlations. These mocks aim to include all
known features in real data. 
A detailed description of the \textsc{EZmocks} can be found in 
\citet{zhaoCompletedSDSSIVExtended2021}. 
These mocks were used in this work to 
estimate covariance matrices and to perform statistical studies of our method.
As we will describe in section~\ref{sec:method:gaussian_approx}, mocks are also required 
by the Gaussian approximation method for combining configuration and Fourier space
results into a single consensus result.

\section{Methodology}
\label{sec:method}

\subsection{Measuring the clustering}

We used correlation function and power spectrum multipoles estimated for each
of the 1000 \textsc{EZmock} realisations. The correlation function is estimated 
using the \citet{landyBiasVarianceAngular1993} estimator, while power spectra 
were calculated using the \citet{yamamotoMeasurementQuadrupolePower2006} estimator, 
as implemented by \citet{bianchiMeasuringLineofsightdependentFourierspace2015, scoccimarroFastEstimatorsRedshiftspace2015}. 

We attempt to remove the impact of non-linear evolution of galaxies and 
a fraction of redshift-space distortions by applying the reconstruction 
technique of \cite{burden_reconstruction_2015}. Using the density field
itself, we estimate the displacements using the Zeldovich approximation
to move the galaxies ``back in time''. 
This technique increases the precision of the measurement by sharpening
the BAO feature. According to \cite{carter_impact_2020} the BAO results 
are not sensitive to small variations in the cosmology used to perform 
the reconstruction. In Appendix \ref{sec:annex:prerecon}, we performed the analysis for the prereconstructed sample, where we derive the same conclusions as for the postreconstruction cases (though at lower signal-to-noise ratios).

\subsection{BAO modelling}
\label{sec:method:bao_model}

The models used in this work are the same employed in \citet{BautistaCompletedSDSSIVExtended2021, GilMarinCompletedSDSSIVExtended2021},
which are themselves based on previous work \citealt{alamClusteringGalaxiesCompleted2017, gil-marinClusteringGalaxiesSDSSIII2016a, gil-marinClusteringSDSSIVExtended2018, rossClusteringGalaxiesCompleted2017, bautistaSDSSIVExtendedBaryon2018}. 
We briefly summarize these models here. 
The code that produces the model and perform the fitting to the data is 
publicly available\footnote{\url{https://github.com/julianbautista/baopy}}.

The aim is to model both the power spectrum $P_\ell(k)$ and correlation function 
multipoles $\xi_\ell(r)$ as a function of wave-number $k$ and separations $r$, 
respectively, where $\ell$ is the order of the multipole expansion. We focus on 
scales relevant for the measurement of the baryon acoustic oscillation (BAO) feature,
typically $0 < k < 0.3~h~{\rm Mpc}^{-1}$ for power-spectrum and $30<r<180~h^{-1}$Mpc for 
correlation function.

The starting point is a linear-theory-based model for the redshift-space anisotropic galaxy power-spectrum $P(k, \mu)$, 
\begin{multline}
    P(k, \mu_k)  = \frac{b^2 \left[1+\beta(1-S(k))\mu_k^2\right]^2}
{(1+ k^2\mu_k^2\Sigma_s^2/2)} \times \\  
\times \left[ P_{\rm no \ peak}(k) + P_{\rm peak}(k)
e^{-k^2\Sigma_{\rm nl}^2(\mu_k)/2}   \right],
\label{eq:pk2d}
\end{multline}
where $b$ is the linear density bias of the galaxy population, 
$\beta$ is the redshift-space distortions parameter defined as the ratio between 
the growth-rate $f$ and $b$,
$k$ is the modulus of the wave-vector $\vec{k}$ and $\mu_k$
is the cosine of the angle between the wave-vector and the
line of sight. 
The broadening of the BAO peak caused by non-linear clustering is 
reproduced by applying an anisotropic Gaussian smoothing, that is 
by multiplying the ``peak-only'' power spectrum 
$P_{\rm  peak}$ (see below) by a Gaussian function with dispersion given by 
$\Sigma_{\rm nl}^2(\mu_k) = \Sigma_\parallel^2 \mu_k^2 + \Sigma^2_\perp(1-\mu_k^2)$.
The non-linear random motions on small scales are modeled by a 
Lorentzian distribution parametrized by $\Sigma_s$.

Following \citet{seoModelingReconstructedBAO2016}, the term 
$S(k) = e^{-k^2\Sigma_r^2/2}$ for the postreconstruction model and 
$S(k)=0$ for the prereconstruction, where $\Sigma_r = 15$\hmpc\ 
is the smoothing parameter used when reconstructing the galaxy catalogues. 

We follow the procedure from \cite{kirkbyFittingMethodsBaryon2013} 
to decompose the BAO peak component $P_{\rm peak}$ from the full linear 
power-spectrum $P_{\rm lin}$. We start by computing the correlation 
function by Fourier transforming $P_{\rm lin}$, then we replace the 
correlations over the peak region by a polynomial function fitted using 
information outside the peak region ($50 < r < 80$ and $160 < r < 190$\hmpc). 
The resulting correlation function is then Fourier transformed back
to get $P_{\rm no \ peak}$.
The linear power spectrum $P_{\rm lin}$ is obtained from  
CAMB\footnote{\url{camb.info}} \citep{lewisEfficientComputationCosmic2000} 
with cosmological parameters of our fiducial cosmology defined in table \ref{tab:fid_cosmo}.

\begin{table}
\caption{Fiducial cosmologies used in this work. All models are parameterized 
by their fraction of the total energy density in form of total matter
$\Omega_{\rm m}$, cold dark matter $\Omega_{\rm cdm}$, 
baryons $\Omega_{\rm b}$, and neutrinos $\Omega_{\nu}$, 
the Hubble constant $h = H_0/(100 {\rm km} \ {\rm s}^{-1}{\rm Mpc}^{-1})$, 
the primordial spectral index $n_s$ and primordial amplitude of power
spectrum $A_s$ of scalar perturbations. With these parameters we compute the 
normalisation of the linear power spectrum $\sigma_8$ at $z = 0$ 
and the comoving
sound horizon scale at drag epoch $r_{\rm d}$.}
\centering
\begin{tabular}{ccc}
\hline
\hline
  & 
  Baseline &
  EZmocks \\
  \hline

$\Omega_{\rm m}$ & 0.310& 0.307 \\
$\Omega_{\rm cdm}$ &  0.260& 0.259 \\
$\Omega_{\rm b}$ & 0.048 &0.048\\
$\Omega_{\nu}$ & 0.0014 &0.0 \\
$h$ & 0.676& 0.678 \\
$n_s$ & 0.970& 0.961 \\
$A_s[10^{-9}]$ & 2.041& 2.116 \\
$\sigma_8(z=0)$ &0.800& 0.823 \\
$r_{\rm d}$ [Mpc] &147.78 &147.66 \\

\hline
\hline
\end{tabular}
\label{tab:fid_cosmo}
\end{table}

The multipoles of the power-spectrum $P_\ell(k)$ are obtained by integrating 
over $\mu_k$ weighted by the Lengendre polynomials $L_\ell(\mu_k)$:
\begin{equation} \label{eq:pkmultipoles}
P_\ell(k) = \frac{2\ell+1}{2} \int_{-1}^{1} P(k, \mu) L_\ell(\mu)
~{\rm d}\mu ,
\end{equation}
The correlation function multiples $\xi_\ell(r)$ are obtained by Hankel 
transforming the $P_\ell(k)$:
\begin{equation}
\xi_\ell(r) = \frac{i^\ell}{2\pi^2}\int_0^\infty k^2 j_\ell(kr)
P_\ell(k) ~ {\rm d}k ,
\end{equation}
where $j_\ell$ are the spherical Bessel functions. 
These transforms are computed using \textsc{Hankl}\footnote{\url{https://hankl.readthedocs.io/en/latest/}} 
that implements the FFTLog algorithm by 
\citet{hamiltonUncorrelatedModesNonlinear2000}.

The BAO peak position is parametrised via two dilation parameters,
one scaling separations across the line of sight, $\aperp$, 
and one scaling separations along the line of sight, $\apara$. 
The observed $\vec{k}$ is related to the true $\vec{k}^\prime$ by 
$k_\parallel = k^{\prime}_\parallel/\apara$ and $k_\perp = k^{\prime}/\aperp$. 
Therefore, the observed power spectrum relates to the true power spectrum as
\begin{equation}
    P(k_\parallel, k_\perp) = \frac{1}{\aperp^2 \apara}   P^\prime\left( \frac{k_\parallel}{\apara}, \frac{k_\perp}{\aperp} \right) .
\end{equation}
We note that this volume scaling does not apply for the correlation function : \begin{equation}
    \xi(r_\parallel, r_\perp) = \xi'\left(\apara r_\parallel, \aperp r_\perp\right).
\end{equation}

These BAO dilation parameters are related, respectively, to the comoving angular 
diameter distance,
$D_M = (1+z)D_A(z)$, and to the Hubble distance, $D_H = c/H(z)$, by
\begin{equation}
\aperp = \frac{D_M(z_{\rm eff})/r_d}{D_M^{\rm fid}(z_{\rm eff})/ r_d^{\rm fid}} ,
\label{eq:aperp}
\end{equation}
\begin{equation} 
\apara =  \frac{D_H(z_{\rm eff})/r_d}{D_H^{\rm fid}(z_{\rm eff})/ r_d^{\rm fid}} ,
\label{eq:apara}
\end{equation}
where $r_d$ is the comoving sound horizon at drag epoch, and $z_{\rm eff}$ is the
effective redshift of the survey. The $r_d$ rescaling is here to account for the choice of template cosmology used to compute the fixed linear power spectrum. For simplicity, the template cosmology is chosen to match the fiducial cosmology used to estimate distances from redshifts.

We apply the scaling factors exclusively to the peak 
component of the power spectrum $P_{\rm peak}(k)$, effectively removing 
any dependency of these parameters on the smooth part $P_{\rm no~peak}(k)$ 
\citep{kirkbyFittingMethodsBaryon2013}.

In order to properly marginalize over any mis-modelling of the smooth part of 
$P_\ell(k)$ and $\xi_\ell(r)$, we add to our model a linear combination of 
smooth functions of scale, with free amplitudes to be marginalised over. 
These smooth functions can also account for potential unknown
systematic correlations that contaminating our measurements. 
Furthermore, since there are no accurate analytical models for 
correlations postreconstruction 
(the $S(k)$ term in Eq.~\ref{eq:pk2d} is generally not sufficient),
these smooth functions can also account for this mis-modelling. As these smooth functions are highly correlated with the finger of god Lorentzian, we fix the parameter $\Sigma_s$ to zero.
Our final template can be written as:
\begin{equation}
P^t_\ell(k) = P_\ell(\aperp, \apara, k) + 
    \sum_{i=i_{\rm min}}^{i_{\rm max}} a^P_{\ell, i}{k^i},
\label{eq:template_pk}
\end{equation}
\begin{equation}
\xi^t_\ell(r) = \xi_\ell(\aperp, \apara, r) + 
    \sum_{i=i_{\rm min}}^{i_{\rm max}} a^\xi_{\ell, i}{r^i},
\label{eq:template_xi}
\end{equation}
where $a^P_{\ell, i}$ and $a^\xi_{\ell, i}$ are the amplitudes for each 
power $i$ of scale, $k$ or $r$, and multipole order $\ell$.
In \citet{GilMarinCompletedSDSSIVExtended2021}, the BAO analysis in Fourier
space used $(i_{\rm min}, i_{\rm max}) = (-1, 1)$, while in \citet{BautistaCompletedSDSSIVExtended2021} the configuration space analysis 
used $(i_{\rm min}, i_{\rm max}) = (-2, 0)$. For both, this corresponds to 
three free parameters per multipole.  
In this work, our baseline choice is $(i_{\rm min}, i_{\rm max}) = (-2, 1)$ 
when performing joint fits unless stated otherwise. 

Our baseline BAO analysis uses the monopoles $P_0, \xi_0$ 
and quadrupoles $P_2, \xi_2$ of the power spectrum and correlation functions (see  Appendix \ref{sec:annex:hexa} for results using the hexadecapole). We fix $\beta = 0.35$ and fitting $b$ with a flat prior between $b=1$ and 4. For all fits, the broadband parameters are free, while both dilation 
parameters are allowed to vary between 0.5 and 1.5.

\subsection{Parameter inference}
\label{sec:method:inference}

The cosmological parameter inference is performed by means of the likelihood analysis of the data. The likelihood $\mathcal{L}$ is defined such that
\begin{equation}
-2\ln\mathcal{L}(\theta) =  \sum_{i,j}^{N_p}\Delta_i(\theta) \hat{\Psi}_{ij} \Delta_j(\theta) + C = \chi^2 + C ,
\label{eq:log_likelihood}
\end{equation}
where $\theta$ is the vector of parameters, $\vec{\Delta}$ is a vector containing
residuals between observed multipoles and their model, $N_p$ is the total number 
of elements in $\vec{\Delta}$. 
An estimate of the precision matrix $\hat{\Psi} = (1-D) \hat{C}^{-1}$ is 
obtained from the unbiased estimate of the covariance from 1000 realisation of EZmocks: $\hat{C}_{ij} = (N_{mock}-1)^{-1} \sum_{k}^{N_{mock}}(X_i^k - \langle X_i\rangle)(X_j^k - \langle X_j\rangle)$ 
where $X_i^k$ is the measured $P_\ell(k_i)$ or $\xi_\ell(r_i)$ for the $k^{th}$ realisation. The factor $D = (N_p+1)/(N_{\rm mocks} -1)$ accounts for the skewed nature of the Wishart distribution \citep{hartlapWhyYourModel2007}. 

The best-fit BAO parameters ($\aperp, \apara$) are determined by minimizing
$\chi^2$ of Eq.~\ref{eq:log_likelihood} using a quasi-Newton minimum finder
algorithm {\sc iMinuit}\footnote{\url{https://iminuit.readthedocs.io/}} which marginalizes over nuisance parameters while sampling for the parameters of interest.
The uncertainties in $\apara$ and $\aperp$ are estimated with the \textsc{minos}
function provided by {\sc iMinuit}. This functions computes the intervals 
where $\chi^2$ increases by unity, which corresponds to a 68\% confidence 
interval. Gaussianity is not assumed in this calculation and uncertainties
can be asymmetric with respect to the best-fit value. 
The two-dimensional confidence contours in $(\aperp, \apara)$, 
such as those presented in Figure~\ref{fig:contours_data}, are estimated
using the \textsc{contour} function from {\sc iMinuit}. Similarly to {\sc minos}, 
this function scans $\chi^2$ values in two dimensions, looking for the contours 
yielding a $\Delta \chi^2 = 2.3$ or $5.9$ for 68 and 95 per cent confidence 
levels, respectively.

\subsection{Consensus via Gaussian approximation}
\label{sec:method:gaussian_approx}

In previous work, configuration and Fourier space results were combined 
into a single consensus result using the method presented in 
\cite{sanchezClusteringGalaxiesCompleted2017a}. 

The idea of the method is to generate a consensus result from  
$M$ different measurement vectors $\vec{x}_m$, each containing $p$ elements, 
and their covariance matrices ${\bf C}_{mm}$, each of size $p\times p$. 
For example, we want to combine $M=2$ measurements of the vector 
$\vec{x}_m = [\aperp, \apara]$ containing $p=2$ parameters 
from Fourier ($m=1$) and configuration space ($m=2$), 
each with its own $2\times2$ error matrix ${\bf C}_{mm}$ derived from their posteriors.  
The consensus is a single vector $\vec{x}_c$ with covariance ${\bf C}_c$,
for which the expressions 
assume that the $\chi^2$ between individual measurements is the same as 
the one from the consensus result.
The expression for the combined covariance matrix is
\begin{equation}
    {\bf C}_c \equiv \left(\sum_{m=1}^M \sum_{n=1}^M {\bf C}_{mn}^{-1} \right)^{-1} ,
\label{eq:C_c}
\end{equation}
and the combined data vector is
\begin{equation}
    \vec{x}_c = {\bf C}_c \sum_{m=1}^M \left( \sum_{n=1}^M  {\bf C}^{-1}_{nm} \right) \vec{x}_m ,
\label{eq:x_c}
\end{equation}
where ${\bf C}_{mn}$ is a $p \times p$ block from 
the full covariance matrix between all parameters 
and methods $\mathcal{C}$, containing $pM \times pM$ elements, 
defined as
\begin{equation}
    \mathcal{C} = 
\begin{pmatrix}
{\bf C}_{11} & {\bf C}_{12} & \cdots & {\bf C}_{1M} \\
{\bf C}_{21} & {\bf C}_{22} & \cdots & {\bf C}_{2M} \\
\vdots  & \vdots  & \ddots & \vdots  \\
{\bf C}_{M1} & {\bf C}_{M2} & \cdots & {\bf C}_{MM}
\end{pmatrix}.
\label{eq:full_covariance}
\end{equation}
The diagonal blocks ${\bf C}_{mm}$ come from each measurement method $M$,
therefore, assuming Gaussian likelihoods for the $p$ parameters. 
The off-diagonal blocks ${\bf C}_{mn}$ with $m\neq n$ cannot be in principle 
estimated from the data itself. 
These off-diagonal blocks are commonly built from mock catalogues. 
Using a set of many realisations, one can build $\mathcal{C}$ from 
all the realisations of $\vec{x}_m$ for each method. 
We obtain the correlation coefficients $\rho^{\rm mocks}_{p_1, p_2, m, n}$, that is 
the covariance $\mathcal{C}$ normalized by its diagonal elements, 
between parameters $p_1$, $p_2$ and methods $m, n$. 
We scale these coefficients by the diagonal errors from the data, to obtain the
final matrix $\mathcal{C}$ for the data. 

It is worth emphasizing that the matrix $\mathcal{C}$ for our data, 
and more specifically its off-diagonal blocks, depend on that
particular realisation of the data in principle. However, the ones derived from 
mock measurements are ensemble averages. 
We account for this fact by scaling the correlations coefficients 
from the mocks in order to match the maximum correlation coefficient
that would be possible with the data 
\citep{rossInformationContentAnisotropic2015}.
For the same parameter $p_1$ measured by two different methods $m$ and $n$,
we assume that the maximum correlation between them is given by 
$\rho_{\rm max} = \sigma_{p1, m}/\sigma_{p1, n}$, 
where $\sigma_p$ is the error of parameter $p$. This number is computed
for the data realisation $\rho_{\rm max}^{\rm data}$ and for the ensemble of mocks
$\rho_{\rm max}^{\rm mocks}$. 
We can write the adjusted correlation coefficients for one particular 
realisation as
\begin{equation}
    \rho^{\rm data}_{p_1, p_1, m, n} = \rho^{\rm mocks}_{p_1, p_1, m, n} 
                \frac{ \rho^{\rm data}_{\rm max} }{ \rho^{\rm mocks}_{\rm max} }.
\label{eq:cov_adjustment_1}
\end{equation}
The equation above accounts for the diagonal terms of the 
off-diagonal block $C_{mn}$. For the off-diagonal terms, we use
\begin{equation}
\rho^{\rm data}_{p_1, p_2, m, n} = 
\frac{1}{4}\left(\rho^{\rm data}_{p_1, p_1, m, n} + \rho^{\rm data}_{p_2, p_2, m, n}\right)
\left(\rho^{\rm data}_{p_1, p_2, m, m} + \rho^{\rm data}_{p_1, p_2, n, n}\right).
\label{eq:cov_adjustment_2}
\end{equation}
This adjustment of correlation coefficients is an addition to the method
proposed by \citet{sanchezClusteringGalaxiesCompleted2017a} and was first implemented 
in \citet{BautistaCompletedSDSSIVExtended2021} to obtain consensus results. 

In summary, the method proposed by \citet{sanchezClusteringGalaxiesCompleted2017a}
can provide consensus results from strongly correlated measurements of the 
same quantities. However, it assumes Gaussian likelihoods and yields 
Gaussian posteriors on the final parameters, which is not a good approximation 
when measurements are noisy. Also, it relies on mock measurements twice: once
to derive the covariance matrix of two-point functions, from which we derive
each parameter vector $\vec{x}_m$; and once again to derive the full 
parameter-method covariance $\mathcal{C}$, using Eqs.~\ref{eq:cov_adjustment_1} 
and \ref{eq:cov_adjustment_2}. We call this method Gaussian approximation (GA). In the following section, we study joint fits in 
Fourier and configuration space, such that mock catalogues are used only 
once to derive a consensus result.


\subsection{Joint analysis}
\label{sec:method:joint}


We studied a new alternative method to get a unique set of constraints on the 
parameters $\alpha_{\parallel}$ and $\alpha_{\perp}$, taking into account 
the full correlations between the configuration space (CS) 
and Fourier space (FS) measurements. 
We refer to this method as Joint Space (JS) analysis. We concatenate the measured $\xi_\ell$ multipoles with the $P_\ell$ multipoles 
such that we obtain a data vector $\left[\xi_0,\xi_2,P_0,P_2\right]$ with 
$\left(2\times r_{bins}+2\times k_{bins}\right)$ data points. 
We then fit the data using a unique set of parameters, except for 
the nuisance parameters defining the polynomial smooth functions 
(Eqs.~\ref{eq:template_pk} and \ref{eq:template_xi}). 

\begin{figure}
	\centering
	\includegraphics[width=1\columnwidth]{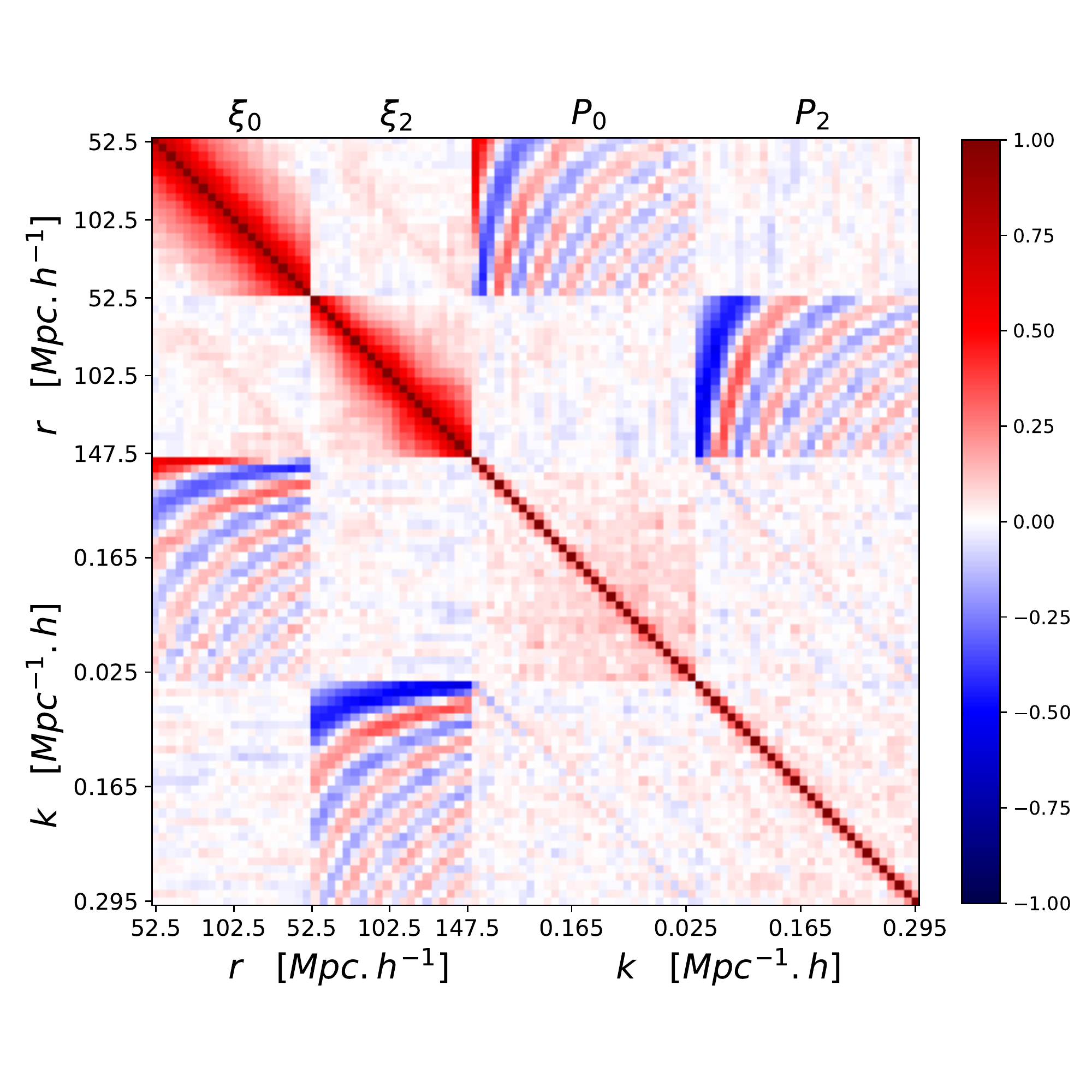}
	\caption{Correlation matrix of the correlation function $\xi_\ell$ and power spectrum multipoles $P_\ell$ estimated using the 1000 independent measurements of EZmocks.}
	\label{fig:covariance}
\end{figure}

In order to perform the joint inference, one needs to estimate the full 
covariance matrix. We use the measured $P_\ell$ and $\xi_\ell$ on the 
1000 EZmocks to compute the covariance matrix $\boldsymbol{\hat{C}}$. 
As in \cite{BautistaCompletedSDSSIVExtended2021} and 
\cite{GilMarinCompletedSDSSIVExtended2021}, we apply the following 
scale cuts: $r \in [50,150] \ h^{-1}{\rm Mpc}$ 
and $k \in [0.02,0.3] \ h{\rm Mpc}^{-1}$. 
Figure \ref{fig:covariance} shows the resulting correlation matrix, 
defined as $\boldsymbol{R} = \boldsymbol{\sigma}^{-1}\boldsymbol{\hat{C}}[\boldsymbol{\sigma}^{-1}]^T$,where $\boldsymbol{\sigma}$ is the vector composed by the square-root  
of the diagonal elements of ${\bf \hat{C}}$. 
The diagonal blocks show the well-known correlations between the monopole 
and quadrupole in the same space, while the off diagonal blocks reveal 
the correlation patterns between the two spaces. The correlations between the two spaces monopole-monopole and quadrupole-quadrupole present the same non-linear features. A given physical scale $r$ is correlated (and anti-correlated) with several $k$ modes, meaning that the statistical information contained in a given scale is spread out over several spectral modes. A physical modeling of these features is proposed in Appendix~\ref{sec:annex:cross_cov}.

\begin{figure}
\centering
  \includegraphics[width=\columnwidth]{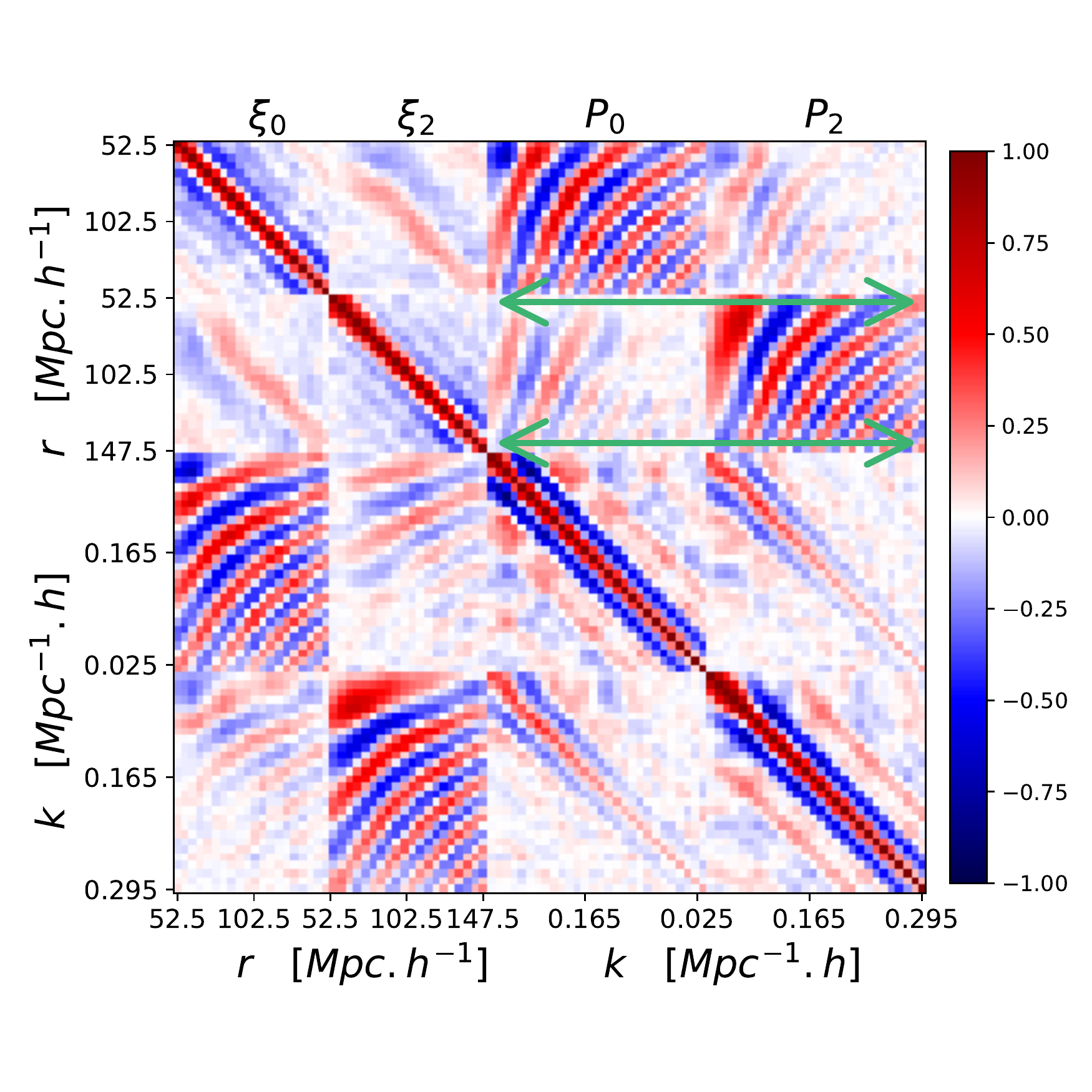}  
  \includegraphics[width=\columnwidth]{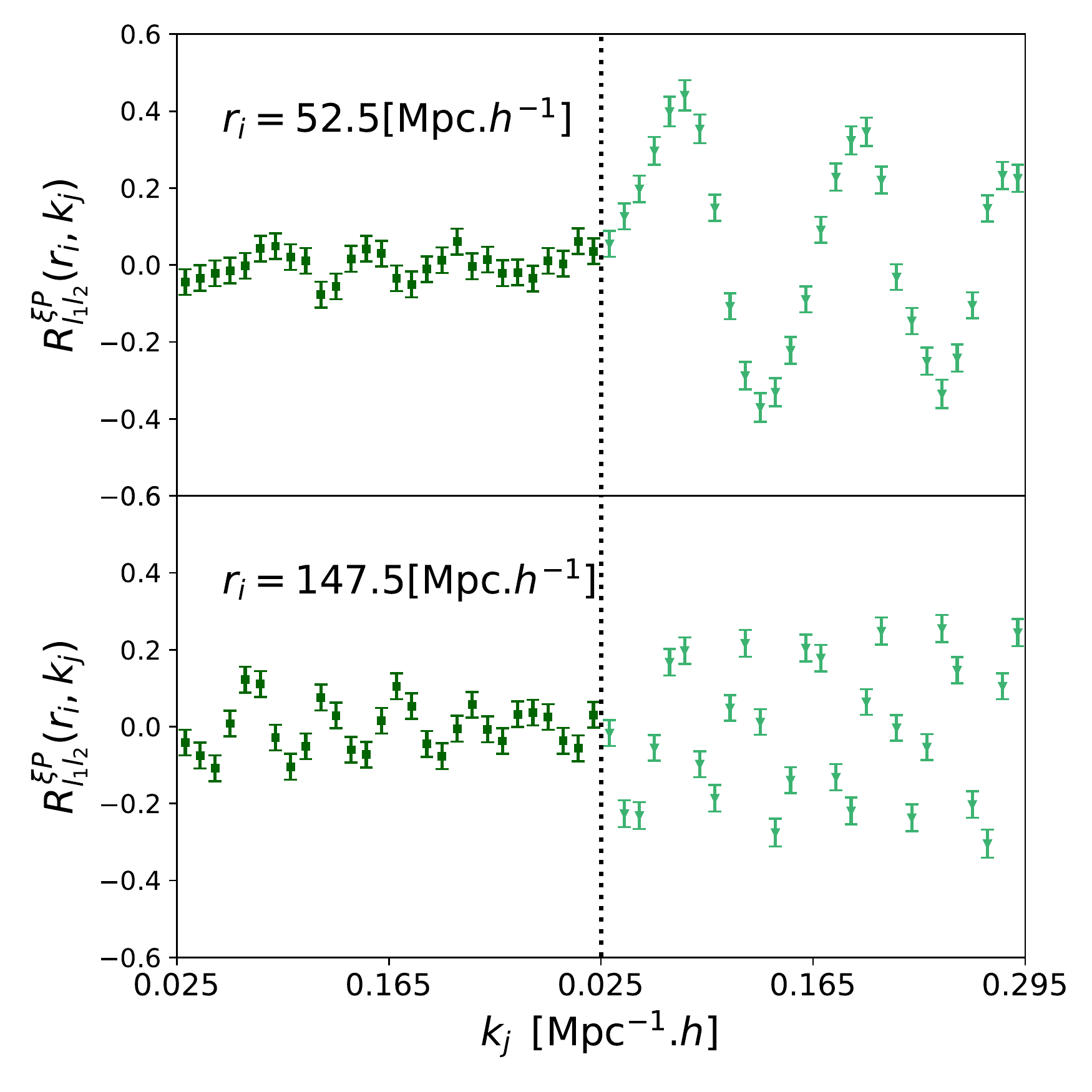}  
\caption{Precision matrix of the correlation function $\xi_\ell$ and power spectrum multipoles $P_\ell$.Top panel: normalized estimated precision matrix 
$R^{\xi P}_{ij} = {\Psi_{ij}}/{\sqrt{\Psi_{ii}\Psi_{jj}}}$, 
obtained from 1000 EZmock realisations. 
	Bottom panel: slices of $R^{\xi P}_{ij}$ showing the cross-correlation between 
	$\xi_\ell$ and $P_\ell$ for particular values of scales as shown
	by the green arrows in the left panel. Errors are estimated from Eq. 29 of \cite{taylorPuttingPrecisionPrecision2013}.
	}
\label{fig:precision_matrix}
\end{figure}


We then need an estimate for the precision matrix 
$\boldsymbol{\hat{\Psi}}$ in Eq.~\ref{eq:log_likelihood}. 
Inverting the covariance mixes the different modes and scales across CS and FS. The top panel of  figure~\ref{fig:precision_matrix} presents the resulting correlations in the 
normalized precision matrix (for convenience we call the coefficients of the precision matrix 'correlations'). Green arrows indicate the regions detailed in the bottom panels, where we focus on the correlations between the CS $\xi_2$ and the FS $P_0$ and $P_2$, respectively by ploting the amplitude of the precision matrix at two fixed scaled $r_{\rm min} = 52.5 \ h^{-1}{\rm Mpc}.$ and $r_{\rm max} = 147.5 \ h^{-1} {\rm Mpc}$. The error bars of the precision matrix are computed according to Eq. 29 of \cite{taylorPuttingPrecisionPrecision2013}. We see that at both scales, $\xi_2$ is weakly correlated with $P_0$ while being strongly correlated with $P_2$. This result is in agreement with the correlations between $\xi_0$ and $\xi_2$. Furthermore, for large $r$, while the number of correlated $k$ modes increases, the amplitude of the correlations decreases distinctly for $\xi_2 - P_2$, and increases for $\xi_2 - P_0$. Note that the correlations between the $\xi_0$ and $P_\ell$ are not shown here but behave in a similar fashion. 

 While capturing additional information about the scale dependence of the joint space correlations, our new methodology has some drawbacks. Indeed, because the covariance matrix $\hat{C}$ we use to infer our set of parameters is an estimate of the real precision matrix drawn from a Wishart distribution (finite sample of mocks), each element is affected by its own uncertainty that should be correctly propagated to the uncertainty of the estimated parameters. It has been shown in \cite{Dodelson2013} that this additional ``noise'' is directly proportional to the parameter covariance. We therefore need to apply correction factors to the obtained $C_\theta$. These factors are given in \cite{Percival2014} as:

\begin{equation}
\begin{split}
    &m_1 = \frac{1+B(N_{\rm bins}-N_{\rm par})}{1+A+B(N_{\rm par}+1)},\\
    &m_2 = \frac{m_1}{1-D},
\end{split}
\label{eq:m1}
\end{equation}
with $D$ the Hartlap factor defined in section\ref{sec:method:inference} and
\begin{equation}
\begin{split}
    &A=\frac{2}{(N_{\rm mock}-N_{\rm bins}-1)(N_{\rm mock}-N_{\rm bins}-4)}, \\
    &B = \frac{(N_{\rm mock}-N_{\rm bins}-2)}{(N_{\rm mock}-N_{\rm bins}-1)(N_{\rm mock}-N_{\rm bins}-4)}.
\end{split}
\label{eq:m1}
\end{equation}

Note that this is also true for a classic FS or CS analysis but since 
the correction factors only scale with the number of parameters, data bins 
and mocks used to estimate the covariance, the enlargement on the constraints 
is smaller. The factor $m_1$ is to be directly applied to the estimated 
parameter covariance matrix for a given measurement, while the factor 
$m_2$ scales the standard deviation of a given parameter over a set of mocks. The values of the parameters $m_1$ and $m_2$ are given in table \ref{tab:correction_factor}. Since the enlargement of the parameter constraints ($\sqrt{m_1}$) expected in CS is about $1\%$, $1.7\%$ in FS and $3\%$ in JS, we expect slightly looser constraints for the joint analysis.

\begin{table}
\caption{Correction factors for the three analysis performed in this work.
$m_1$ is the factor to be applied to the estimated covariance matrix of the parameters and $m_2$ is the factor that scales the scatter of best-fit parameters of a set of mocks (if these were used in the calculation of the covariance matrix). $N_{\rm mock}$ is the number of mocks used in the estimation of the covariance matrix, $N_{\rm par}$ is the total number of parameters fitted and $N_{\rm bins}$ is the total size of the data vector. The derivation of $m_1$ and $m_2$ can be found in \cite{Percival2014}.}
\centering
\begin{tabular}{cccccc}
\hline
\hline
 Analysis & 
  $m_1$ &
  $m_2$ &
  $N_{par}$ &
  $N_{bins}$ &
  $N_{mock}$\\
  \hline
CS & 1.018 & 1.061 & 11 & 40 & 1000 \\ 
FS & 1.035 & 1.097 & 11 & 56 & 1000 \\ 
JS & 1.062 & 1.176 & 19 & 96 & 1000 \\ 

\hline
\hline
\end{tabular}
\label{tab:correction_factor}
\end{table}

\section{Results on mock catalogues}
\label{sec:results_mocks}

In this section, we use postreconstruction \textsc{EZmocks} to validate 
our parameter inference and error estimation (see  Appendix \ref{sec:annex:prerecon} for a discussion on the analysis of the prereconstruction \textsc{EZmocks}). The aim is to compare
results from configuration space (CS), Fourier space (FS), Joint space (JS) with 
the consensus results from the Gaussian approximation (GA).

\subsection{Fits on average correlations}
\label{sec:results_mocks:average} 

Using the inference methodology described in the section~\ref{sec:method}, 
we fit the average $P_\ell$ and $\xi_\ell$ of the 1000 EZmocks in order 
to study potential biases and compare the GA and JS methods. 

At first, we model $\xi_\ell$ and  $P_\ell$ in the separation ranges $r \in [50,150] \ h^{-1}{\rm Mpc}$ and $k \in [0.02,0.3] \ h{\rm Mpc}^{-1}$, by setting the  broad band smooth polynomial functions from Eqs.~\ref{eq:template_pk} and \ref{eq:template_xi} to the shape $\rm (i^{min},i^{max})=(-2,1)$ and letting all fitting parameters free. Then, we assess the robustness of the best fit parameters with respect to variations in $\Sigma_\perp, \Sigma_\parallel$, k ranges, r ranges, number of broadband terms, and template cosmology for the linear power spectrum. Each time we vary one of those settings we keep the other ones fixed in all spaces (CS, FS and JS), facilitating the 
comparisons of results. Since we fit the mean of the mocks, we normalize the covariance matrix by the total number of mocks $N_{\rm mocks} = 1000$. As the covariance is estimated with the same 1000 realisations, we do not expect it to be highly accurate, but sufficient for our purposes.

\begin{figure}
\centering
\includegraphics[width=0.8\columnwidth]{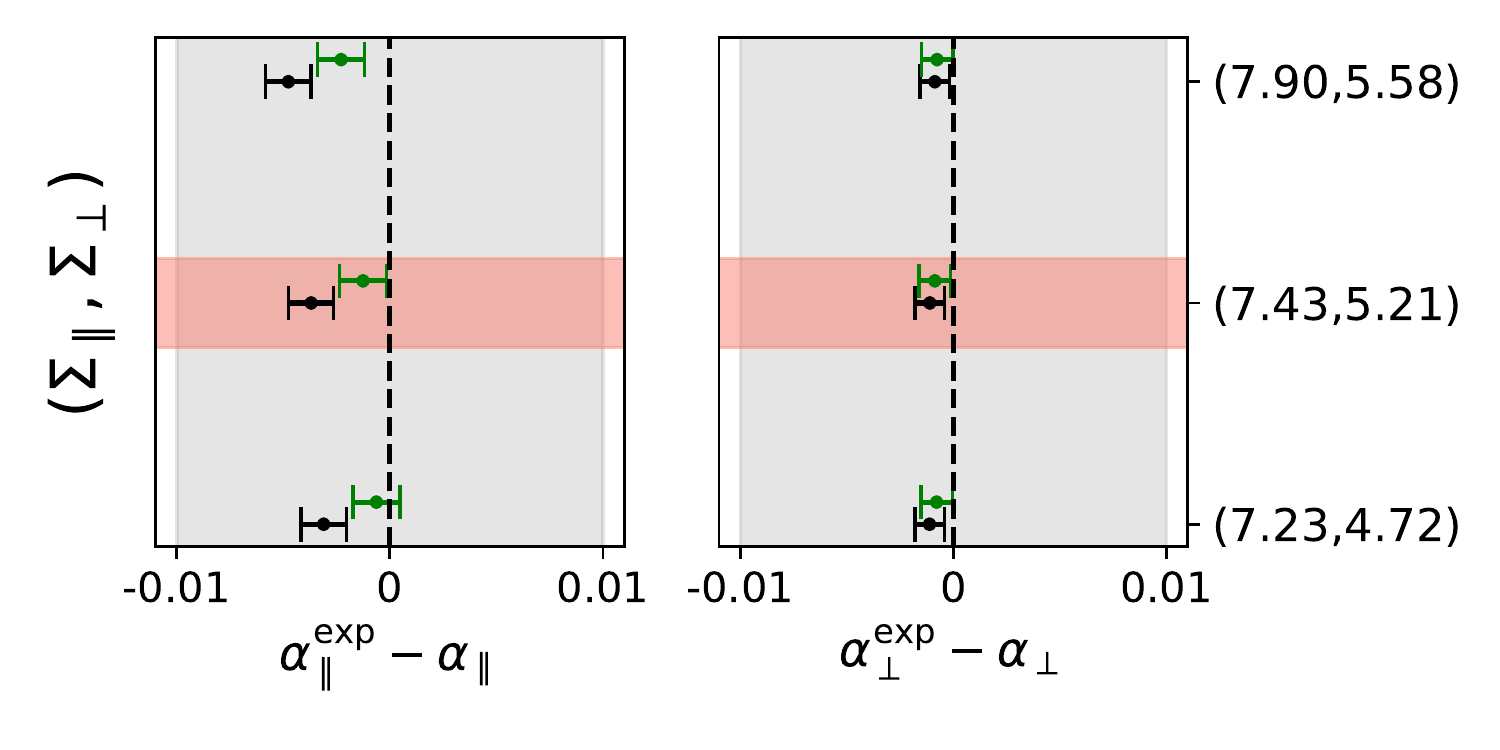} 
\includegraphics[width=0.8\columnwidth]{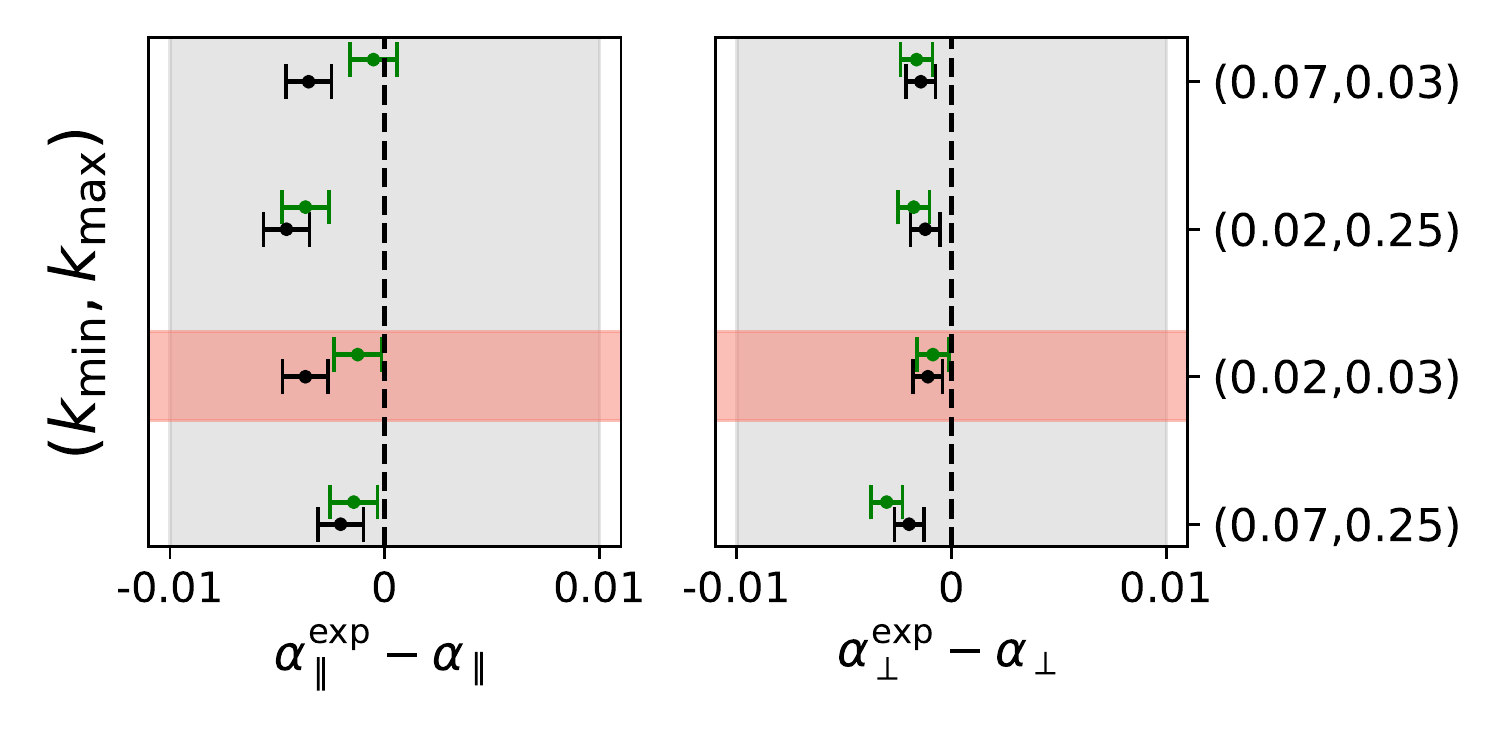}  
\includegraphics[width=0.8\columnwidth]{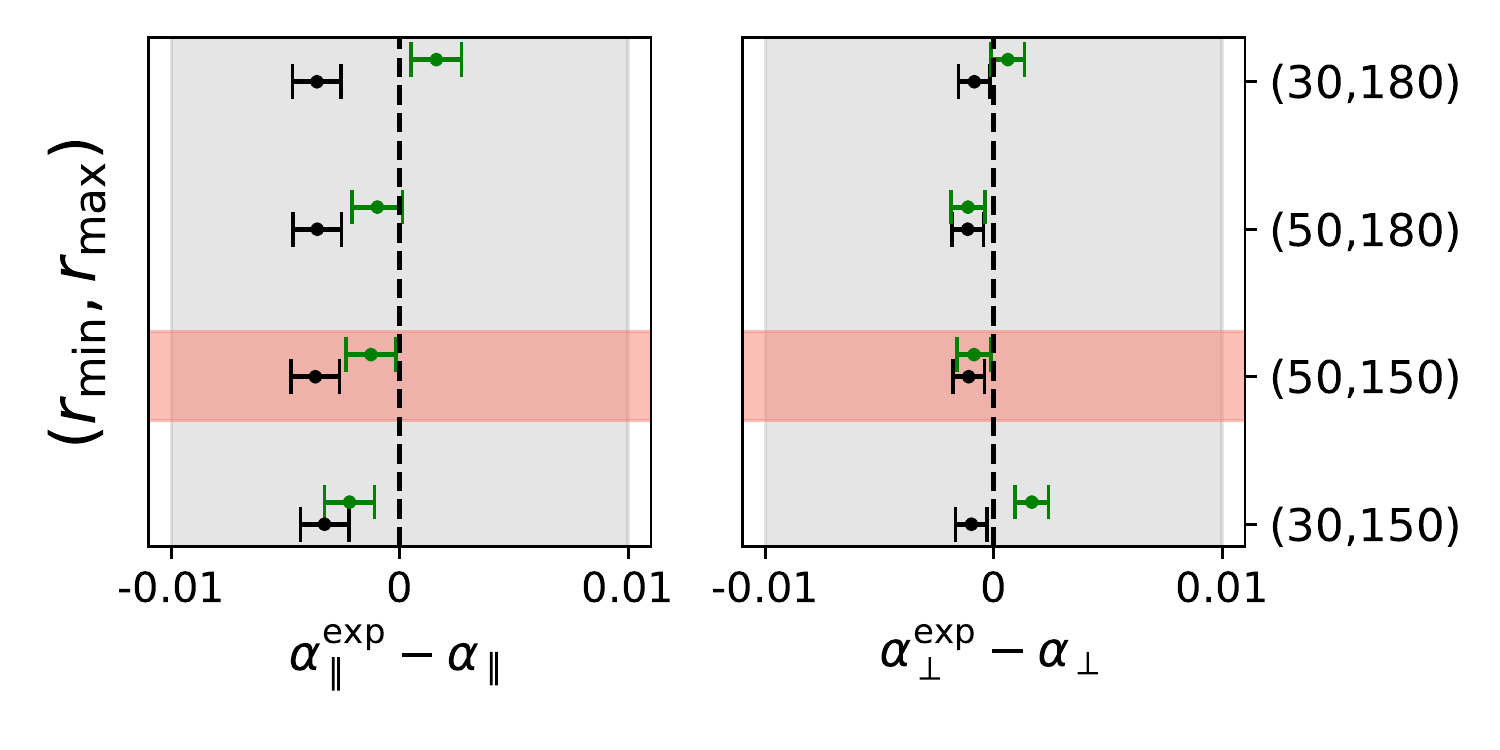}  
\includegraphics[width=0.8\columnwidth]{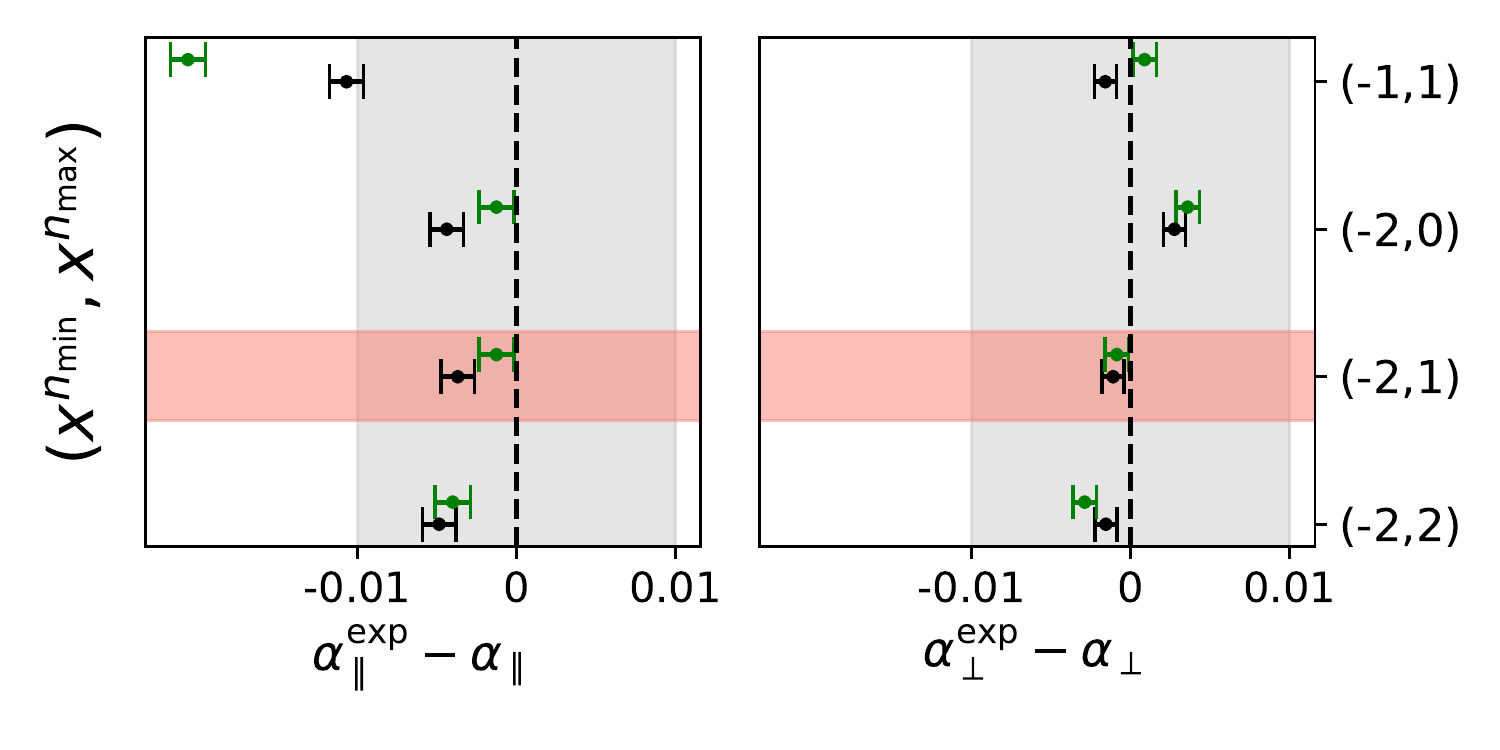}  
\includegraphics[width=0.8\columnwidth]{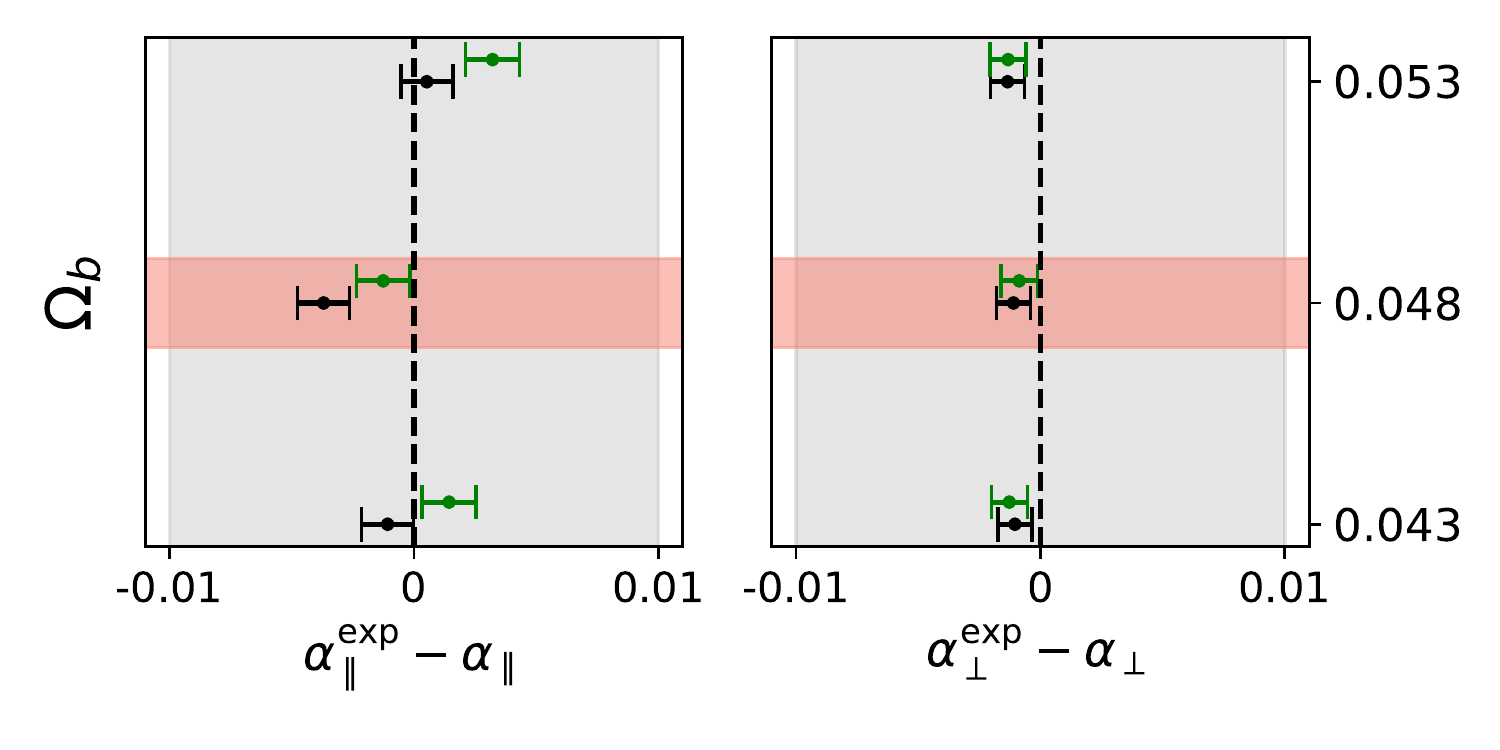}  
\includegraphics[width=0.8\columnwidth]{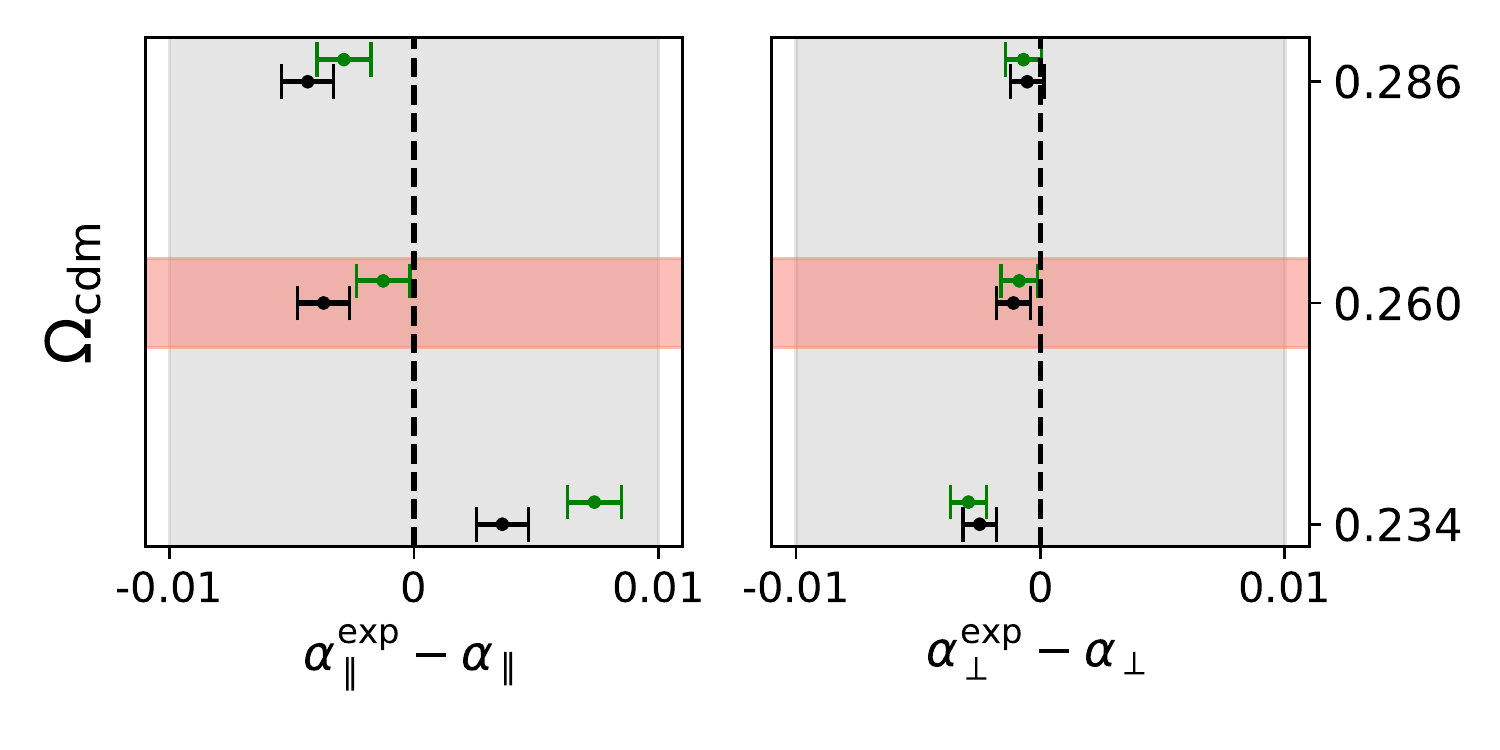}  
\caption{Impact of the choice of fitting scales, fiducial cosmology, broadening $\Sigma$'s parameters and polynomial broadband order in the recovered values of the parameters $\apara$ and $\aperp$. Each point is the best-fit from the average of the 1000 EZmocks. The JS results are in green and GA in black. The grey shaded areas correspond to a 1 percent error and red shared areas indicate the fiducial choices of our analysis.} 
\label{fig:syst_postrecon}
\end{figure}

Figure~\ref{fig:syst_postrecon} presents the impact of different analysis
settings on the best fit values of $\alpha_\parallel$ and $\alpha_\perp$ 
for the GA (in black) and JS (in green) analysis. 
In every plot, a gray shaded area represents the 1 percent deviation 
from the expected value, which was the tolerance used in previous analysis.
First, we tested different set of values for 
the broadening of the BAO peak, $(\Sigma_\parallel,\Sigma_\perp)$, 
corresponding to the best values found in CS (7.90, 5.58), 
FS (7.23, 4.72) and JS (7.43, 5.21), all in units of $h^{-1}$Mpc. 
Both GA and JS analyses are robust to changes in those parameters. 
Both GA and JS analyses are also robust for different ranges of scales
used in the fit, presenting small and 
coherent shifts. 
This indicates that the information is correctly extracted only from 
the BAO feature. 
When varying the broadband terms, the JS method shows stable results except 
for $\apara$ in one setting (-1, 1), where the broadband is not flexible 
enough to fit the residuals. 
When increasing the number of polynomial terms, the systematic shifts are 
notably smaller in JS than GA. 
The last two panels of Figure~\ref{fig:syst_postrecon} show 
the influence of the template cosmology used to compute 
the linear power spectrum (not the one to convert redshifts into distances). 
We chose to vary separately the two parameters
$\Omega_{\rm b}$ and $\Omega_{\rm cdm}$. 
Each time, we compute the new 
expected values for the $\alpha$ parameters by rescaling 
$r_d^{\rm fid}$ in Eqs. \ref{eq:aperp} and \ref{eq:apara} 
(the fiducial distances are unchanged as the cosmology used to 
compute distances does not vary). We find that the systematic shift 
behave in the same way for the two methods, and do not exceed 1$\%$ when varying $\Omega_{\rm cdm}$ or $\Omega_{\rm b}$ by 10$\%$ 
from the baseline cosmology. Note however that we performed the fits while letting free the parameters $\Sigma_\parallel$ and $\Sigma_\perp$, which helps reducing the shift amplitude between the different cosmologies.

Overall,  the inferred value for the parameter 
$\alpha_\perp$ is the more robust than $\apara$ for both methods. 
Globally, GA and JS behave in the same way when varying the fitting
configuration, proving that these are not caused by the new methodology. 
Best values and uncertainties are consistent between GA and JS. 
The systematic shifts for $\alpha_\parallel$ are smaller in JS, while 
no significant difference is seen for $\aperp$.
The chosen baseline for the rest of the analysis is highlighted in red. 
The systematic errors may partially result from the covariance matrix that would require more realisations, but as they are much smaller than statistical errors budget for the real data ($\sigma_\alpha /\alpha \sim 2 \%$), we 
can safely neglect these.

\subsection{Fits on individual mocks}
\label{sec:results_mocks:individual} 

We now focus on the statistical properties of best-fit values and 
their uncertainties, 
based on fits of individual realisations of EZmocks. 
When fitting individual mocks, we fix 
$\left( \Sigma_\parallel, \Sigma_\perp \right)$
to the best-fit values of the Joint analysis in the stack of 1000 measurements 
(see previous section), while letting all other parameters free.  
Table~\ref{tab:priors} summarises the parameters and flat priors used in these fits.
The constraints on  $\left( \alpha_\parallel, \alpha_\perp \right)$ are 
then obtained by marginalizing over $b$ and all other nuisance parameters. 

For each method (FS, CS and JS) we remove results from nonconverging
likelihoods (unsuccessful contour estimation) and extreme best-fit values at 
5$\sigma$ level ($\sigma$ is defined as half of the range covered by 
the central 68 per cent values). The best-fit values for which the 
estimated uncertainties touch the prior boundaries are removed as well. The remaining number of realisations referred as $N_{\rm good}$ is given for each method in table \ref{tab:stats} and is consistent with the results of \cite{GilMarinCompletedSDSSIVExtended2021} and \cite{BautistaCompletedSDSSIVExtended2021}.

\begin{table}
    \centering
    \caption{Parameters used in the fits, with their flat priors. 
    See Eq.~\ref{eq:pk2d} and text for detailed description of the parameters.
    }
    \begin{tabular}{ccc}
    \hline 
    \hline 
        Parameter & Prior & Is fixed ?  \\
    \hline
        $\aperp$ & [0.5, 1.5] & No \\
        $\apara$ & [0.5, 1.5] & No \\
        $b$      & [1.,  4.] & No \\
        $\beta$   &  -       & 0.35 \\ 
        $\Sigma_\parallel $ &   - & 7.428 \\
        $\Sigma_\perp $ &   - & 5.210 \\
        $a^P_{\ell, i}$ & - & No \\
        $a^\xi_{\ell, i}$ & - & No \\
    \hline 
    \hline 
    \end{tabular}
    \label{tab:priors}
\end{table}

\begin{figure}
	\centering
	\includegraphics[width=0.8\columnwidth]{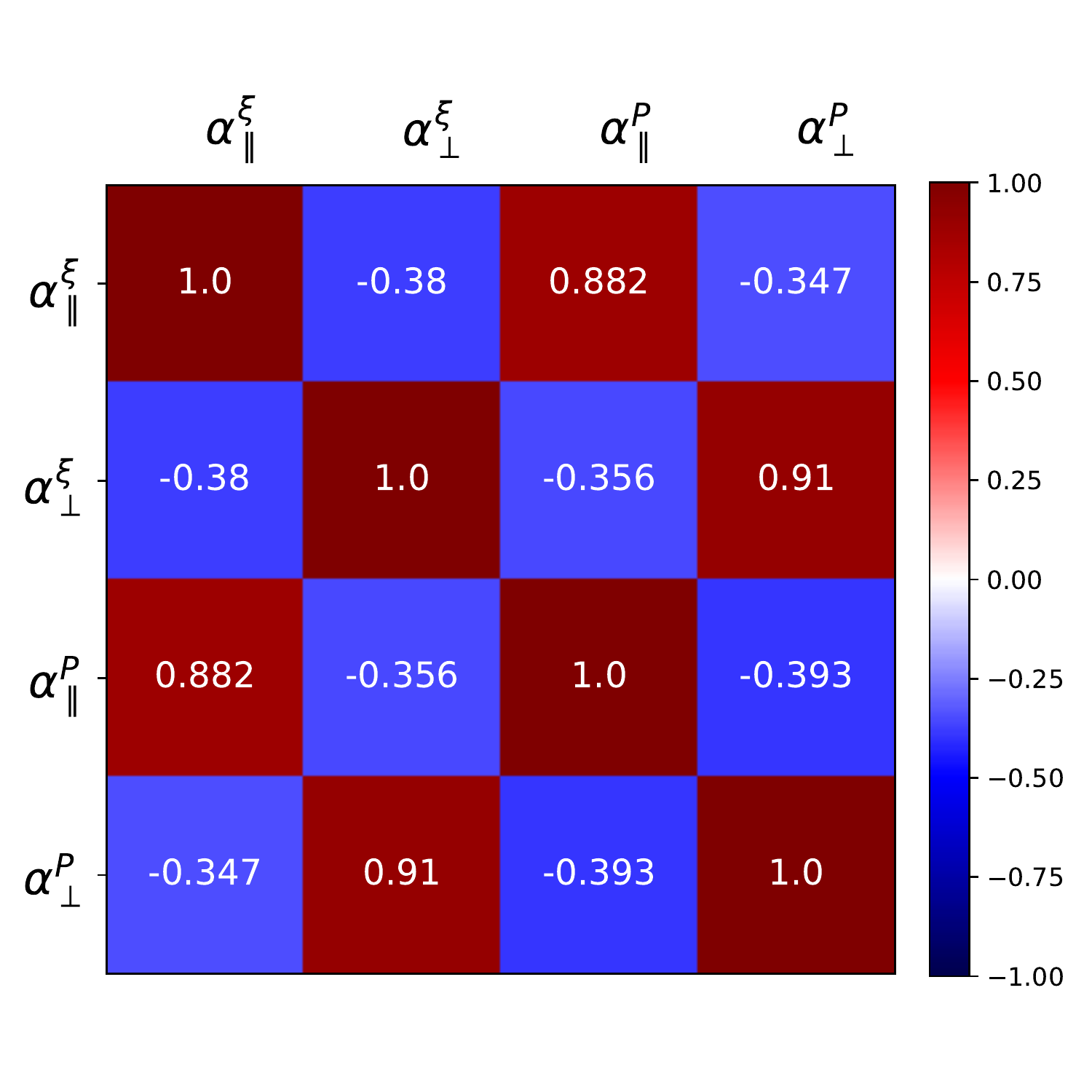}
	\caption{Correlation coefficients between $\alpha_\parallel$ and $\alpha_\perp$ in CS and FS obtained from fits to the 1000 EZmock realisation.}
	\label{fig:alpha_corr_tot}
\end{figure}

To obtain GA results, we extract for each mock posterior profile 
the 1 sigma contour of the parameter space. 
We fit the likelihood contours corresponding to 68 per cent C.L. 
with an ellipse, which we translate into a parameter covariance matrix
$C_{mm}$, where $m$ refers to either CS or FS.
We scale the resulting covariance with the parameter $m_1$ 
(see table~\ref{tab:correction_factor}).
We then construct the total covariance matrix $\mathcal{C}$ from 
Eq. \ref{eq:full_covariance} obtained from the 1000 best-fit $(\aperp, \apara)$,
adjusting each time the coefficients (according to Eqs. \ref{eq:cov_adjustment_1} and 
\ref{eq:cov_adjustment_2}) to account for the observed errors of a 
given realisation. The corresponding correlation matrix before the 
individual adjustments is shown in Figure \ref{fig:alpha_corr_tot}. 
Using the combination method, we compute for each mock 
the consensus data vector $\vec{x}_c$ and covariance $C_c$ from Eqs.~\ref{eq:C_c}
and \ref{eq:x_c}. Here we emphasize that the strong assumption of 
Gaussian elliptic contours for the parameters 
$\left(\alpha_\parallel, \alpha_\perp \right)$ of every 
inference highly depends on the statistic of the considered tracer. 
The presence of non-Gaussian contours in our sample of mocks introduces
biases in the total covariance matrix $\mathcal{C}$ and each individual 
consensus data vector $\vec{x}_c$ and covariance ${\bf C}_c$. 
This is one of the limitations of the GA method that we can avoid with
a JS fit.

\begin{figure}
\centering 
  Gaussian Approximation 
  \includegraphics[width=\columnwidth]{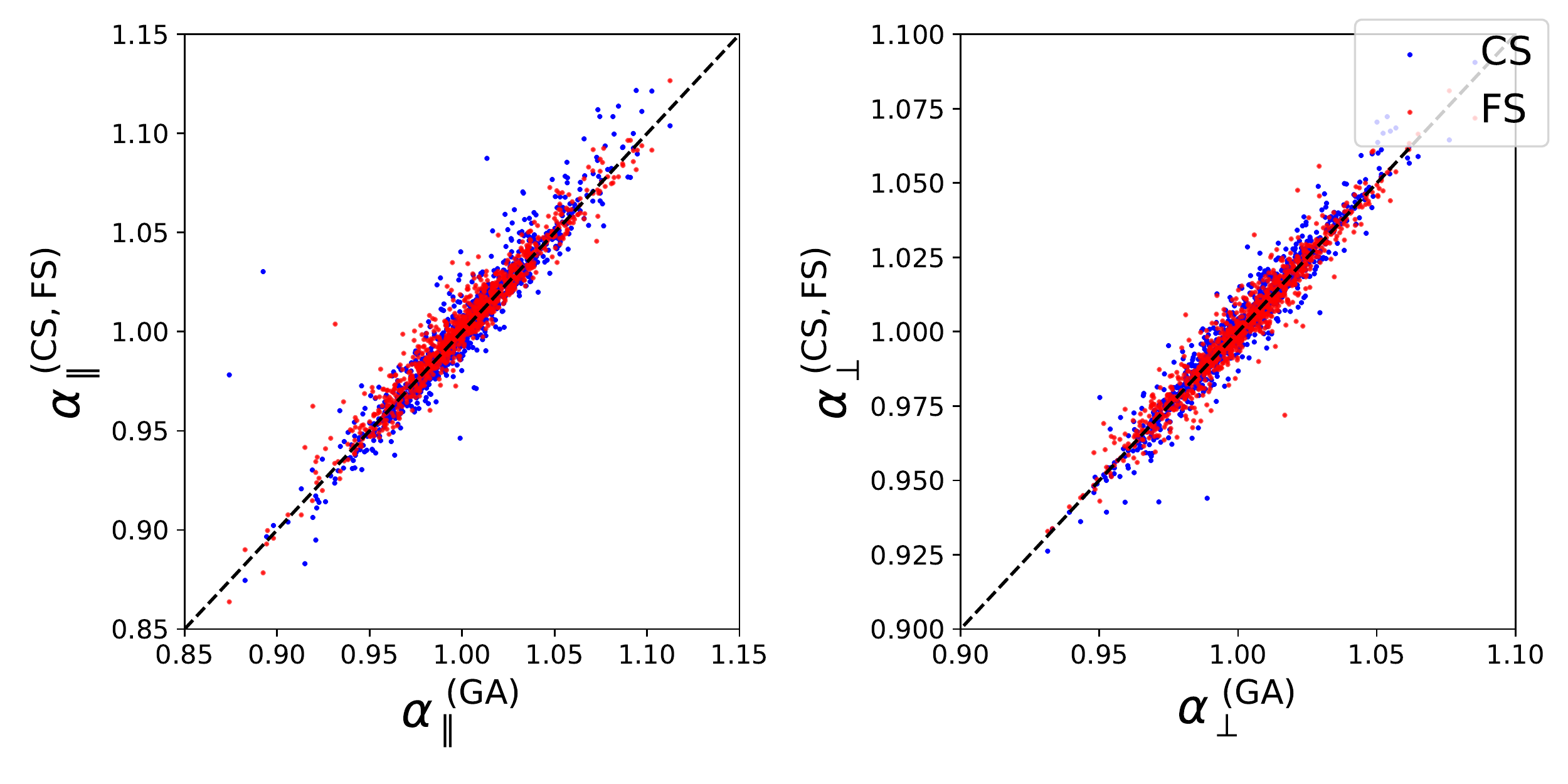}  
  \includegraphics[width=\columnwidth]{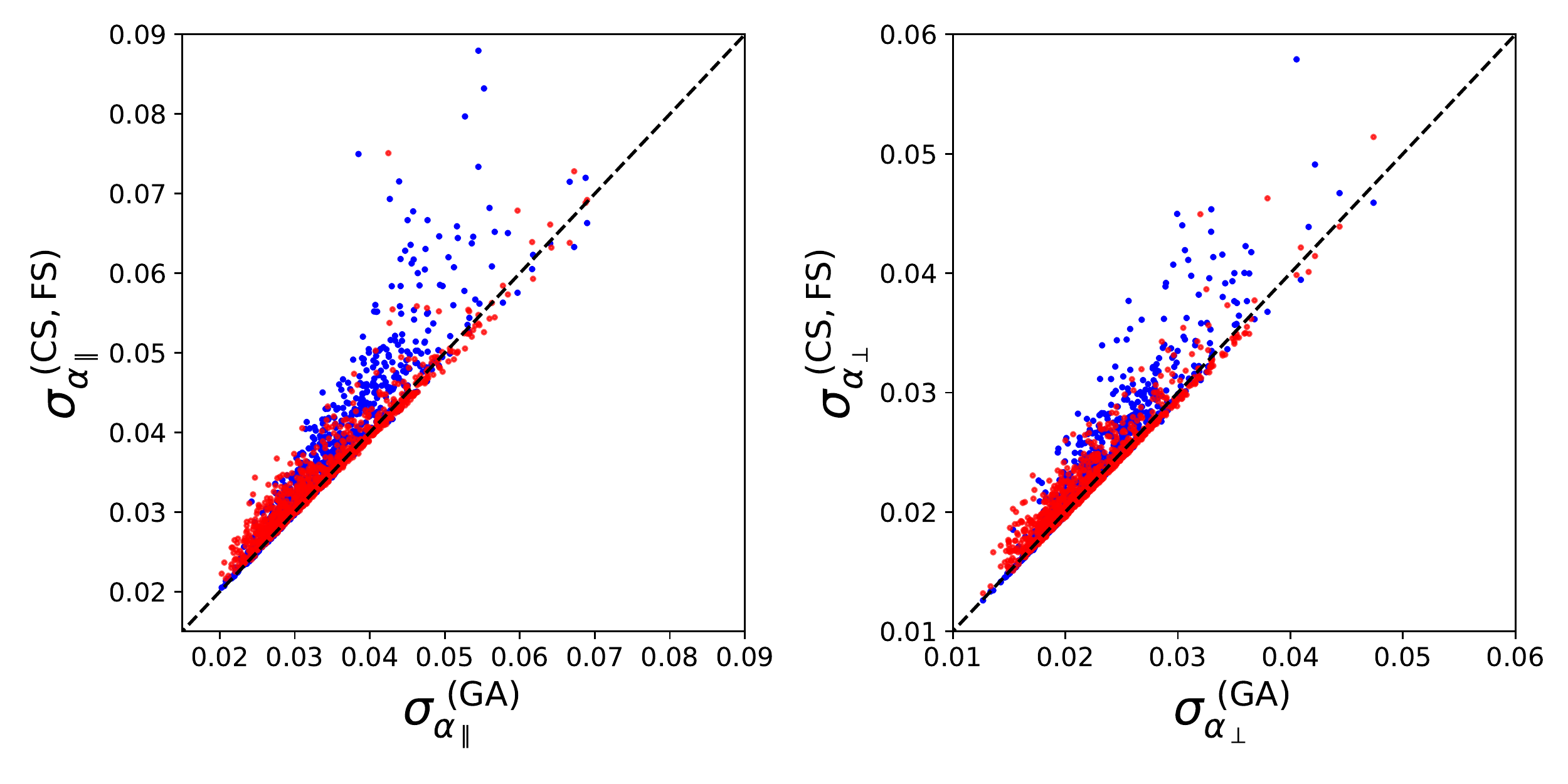} 
  Joint Space fits
  \includegraphics[width=\columnwidth]{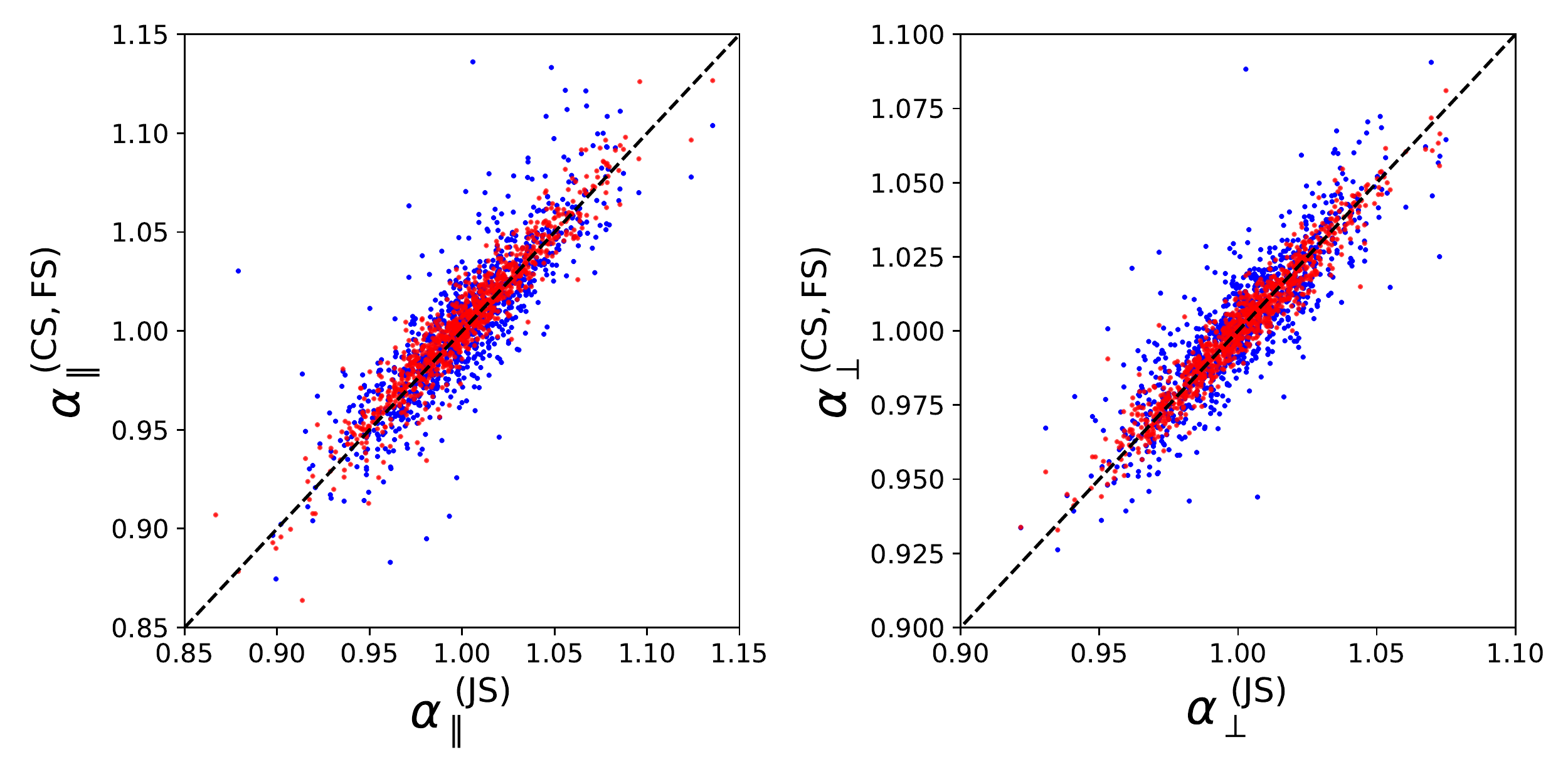}  \includegraphics[width=\columnwidth]{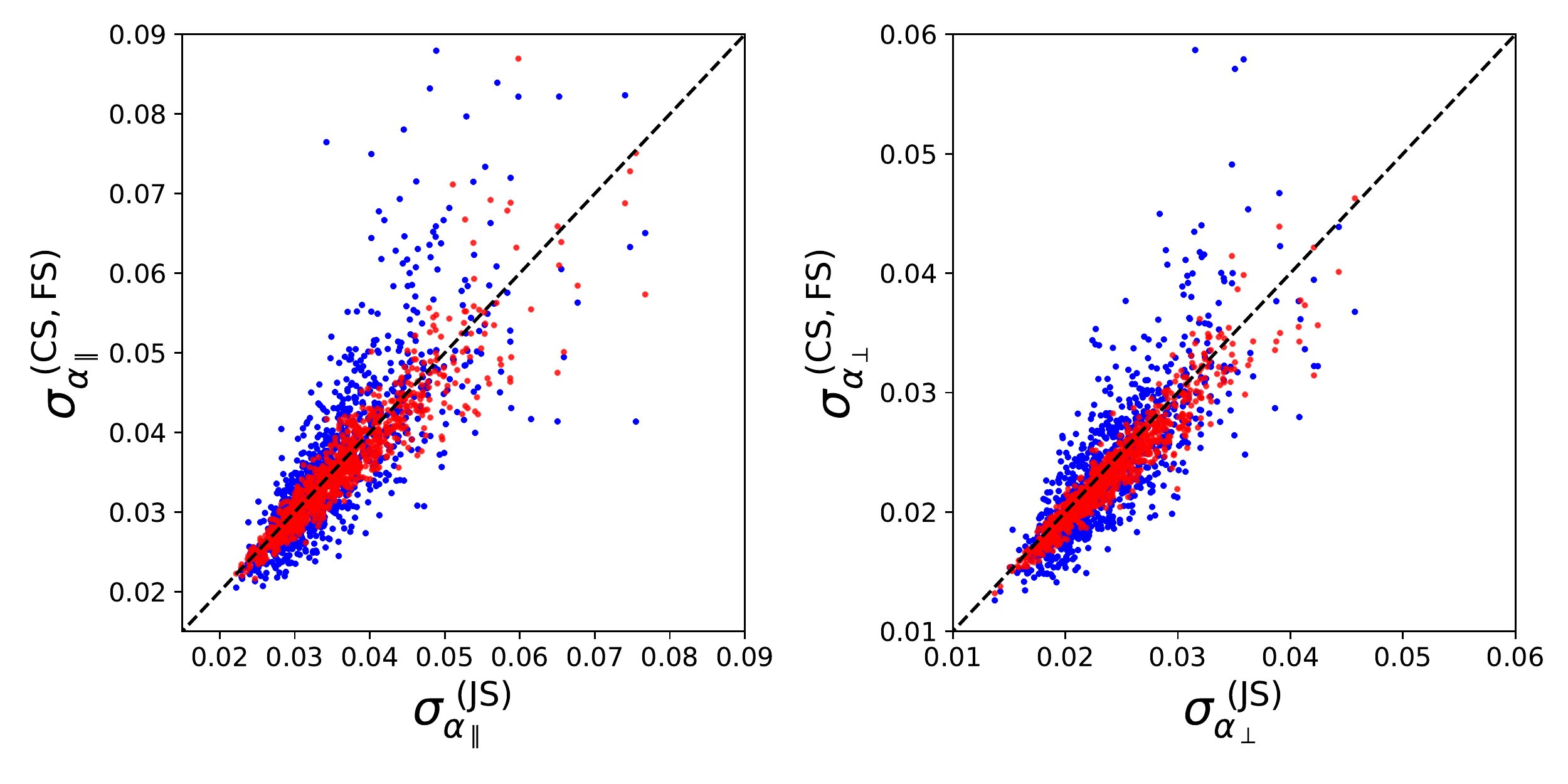}  
  
	\caption{Comparison between the distributions of the best-fit BAO parameters $\left(\alpha_{\parallel},\alpha_{\perp}\right)$ and their estimated errors obtained by fitting individual mock realisations. The four top (bottom) panels compare the CS (blue), FS (red) distributions with the GA (JS). The dashed lines represent a perfect correlation.}
	\label{fig:1000_ezmocks_fit}
\end{figure}

Figure \ref{fig:1000_ezmocks_fit} compares the 
distributions of $\alpha_{\parallel}$ (left)  and $\alpha_{\perp}$ (right) and and their 1$\sigma$ uncertainties as estimated by GA (top four panels) and JS (bottom four panels) versus the same quantities obtained from CS and FS fits. 
We see that the distributions are nicely correlated and scattered 
around the identity line (dashed). 
We find that the scatter (especially for errors)
is less important between FS and JS than CS and JS. 
This would indicate that the Fourier space information has slightly 
more weight than the configuration space in the minimization 
of the likelihood. Moreover, while the GA errors are almost 
systematically inferior to the CS and FS ones, the JS errors are 
scattered on both sides of the dashed lines. 
Due to statistical fluctuations, JS analysis may result in looser constraints
than FS or CS.
We believe these fluctuations might origin from the finite
number of mock realisations used to build the covariance matrix,
though it is hard to test this hypothesis without a larger number 
of mocks.

\begin{table*}
    \centering
    \caption{Statistics on the fit of the 1000 EZmocks realisations. $N_{\rm good}$ is the number of valid realisations after removing undefined contours and extreme values and errors. We show the mean value of the best-fit reduced ${}_r\chi^2_{\rm min}$. For each parameter, we show the average bias $\Delta_\alpha \equiv \langle \alpha_i - \alpha_{\rm exp}\rangle$, the standard deviation of best-fit values $\sigma \equiv \sqrt{\langle \alpha_i^2 \rangle - \langle \alpha_i\rangle^2}$, the average of the per-mock estimated uncertainties $\langle \sigma_i \rangle$, the asymmetry of the estimated error distribution $ A\left(\sigma_i\right) \equiv \langle 2\left(\sigma_{\rm i}^{\rm sup}-\sigma_{\rm i}^{\rm inf}\right)/\left(\sigma_{\rm i}^{\rm sup}+\sigma_{\rm i}^{\rm inf}\right)\rangle $, where $\sigma_i^{sup}$ and $\sigma_i^{\rm inf}$ are the superior and inferior one sigma errors estimated from the likelihood profile, the average of the pull $Z_i \equiv (\alpha_i - \langle \alpha_i \rangle)/\sigma_i$ and its standard deviation $\sigma(Z_i)$. }
    {
    \begin{tabular}{lcc|cccccc|cccccc}
    \hline
    \hline
    & & & \multicolumn{6}{c|}{$\alpha_{\perp}$}  & \multicolumn{6}{c}{$\alpha_{\parallel}$} \\
     & $N_{\rm good}$ & $\langle{}_{r}\chi^2_{\rm min}\rangle$ &
    $\Delta_\alpha $ &
    $\sigma $ & 
    $\langle\sigma_i\rangle$ & 
    $\langle A\left(\sigma_i\right)\rangle$ & 
    $\langle Z_i\rangle$ & 
    $\sigma(Z_i)$ & 
    $\Delta_\alpha $ &
    $\sigma $ & 
    $\langle\sigma_i\rangle $ & 
    $\langle A\left(\sigma_i\right)\rangle$ &
    $\langle Z_i \rangle$ & 
    $\sigma(Z_i)$ \\ 
    & & & [$10^{-2}$] & [$10^{-2}$]  & [$10^{-2}$] & [$10^{-2}$] & & 
        & [$10^{-2}$] & [$10^{-2}$] & [$10^{-2}$] & [$10^{-2}$] & & \\
    \hline

    CS & 985 & 0.96& -0.18& 2.38& 2.35& 2.46& -0.030& 0.998& -0.26& 3.71 & 3.65& 4.63& -0.031& 0.961 \\
    FS & 998 & 0.99& -0.13& 2.29& 2.30& 1.95& -0.020& 0.954& -0.40& 3.58 & 3.54& 0.51& 0.004& 0.930 \\
    GA & 983& -& -0.12& 2.19& 2.24& -& -0.025& 0.989& -0.19& 3.36& 3.40& -& -0.005& 0.963 \\
    JS & 994 & 0.95& -0.12& 2.44& 2.39& 2.40& -0.022& 0.933& -0.12& 3.64& 3.61& 0.10& 0.004& 0.909 \\ 

    \hline 
    \hline
    \end{tabular}
    }
    \label{tab:stats}
\end{table*}

Table~\ref{tab:stats} summarizes the statistical properties of 
$\left(\alpha_\parallel, \alpha_\perp \right)$ for the different analyses,
CS, FS, GA and JS, performed on 1000 EZmock realisations. 
For each parameter, we show six quantities: the average bias $\Delta_\alpha = \langle \alpha - \alpha_{\rm exp} \rangle$ with respect to the
expected value (${\aperp}_{\rm , exp}={\apara}_{\rm , exp}=1$), 
the standard deviation of best-fit values $\sigma$, 
the mean estimated error $\langle \sigma_i\rangle$, 
the mean asymmetry of the estimated error distribution 
$\langle A\left(\sigma_i\right)\rangle =\langle 2\left(\sigma_{\rm i}^{\rm sup}-\sigma_{\rm i}^{\rm inf}\right)/\left(\sigma_{\rm i}^{\rm sup}+\sigma_{\rm i}^{\rm inf}\right)\rangle $,
the mean of the pull $Z_i = \left(\alpha_i - \langle \alpha_i\rangle \right)/\sigma_i$ 
and its standard deviation. 
If errors are correctly estimated and follow a Gaussian distribution, 
we expect that $\sigma = \langle\sigma_i\rangle$, $\langle Z\rangle=0$ and 
$\sigma(Z_i)=1$. 
Table \ref{tab:stats} also shows the number $N_{\rm good}$ of valid realisations after 
removing undetermined likelihoods, extreme values and errors, 
along with the mean value of the reduced chi-square ${}_r\chi^2_{\rm min}$.

Firstly, one can notice that the number of valid realisations (as defined above) 
for the JS analysis is larger than the one of the combined analysis GA. 
The GA method requires both FS and CS fits to converge, 
so the number of valid realisations for GA is necessarily the 
intersection of valid FS and CS realisations. By joining the information of both spaces, the JS fit is able to constrain 
the acoustic scale even on noisy mocks, where FS and CS separately fail.
For each fitting procedure, we find the minimum reduced chi-square  
$\langle{}_{r}\chi^2_{\rm min}\rangle \sim 1$, showing that the majority of the mocks are accurately modeled by our templates.

Secondly, we see good agreement between $\langle\sigma_i\rangle$ and $\sigma$ 
for both parameters in all analyses. While the JS errors lies in between 
the CS and the FS errors for $\alpha_\parallel$, the combined results present 
in average smaller errors. Note that the standard deviation $\sigma$ 
of the best-fit values are scaled with the appropriate correction factor
$\sqrt{m_2}$ (see table \ref{tab:correction_factor}). A crucial 
issue with the GA method is $m_2$ is ill-defined. 

Overall, systematic shifts $\Delta_\alpha$ for all methods are below half a per cent. The GA and JS methods yield reduced systematic shifts $\Delta_\alpha$ than
each individual space alone.
 While the shifts between GA and JS are the same for 
 $\alpha_\perp$, the JS method yields slightly smaller shift for 
 $\alpha_\parallel$ as also observed in the previous section. 
 If we assume that $\sigma$ is a good estimate of the standard deviation of $\alpha$,
 then the uncertainty of $\Delta_\alpha$ should be $\sigma_\Delta = \sigma/\sqrt{N_{\rm good}}$. 
 In this case, all systematic shifts are smaller than 3$\sigma_\Delta$ except for $\apara$ in FS, which reaches a 3.5$\sigma_\Delta$ discrepancy. 
 Shifts for GA are (-1.7, -1.8)$\sigma_\Delta$ for $(\aperp, \apara)$ and 
 (-1.5, -1.0)$\sigma_\Delta$  for JS fits. 
 Note that our results are slightly different from those reported in \cite{BautistaCompletedSDSSIVExtended2021} and \cite{GilMarinCompletedSDSSIVExtended2021} 
 due to a few differences in the analyses, such as how we decompose peak and smooth parts of the
 template, the number of polynomial terms, the values of non-linear damping terms.

\begin{figure}
	\centering
	\includegraphics[width=1.\columnwidth]{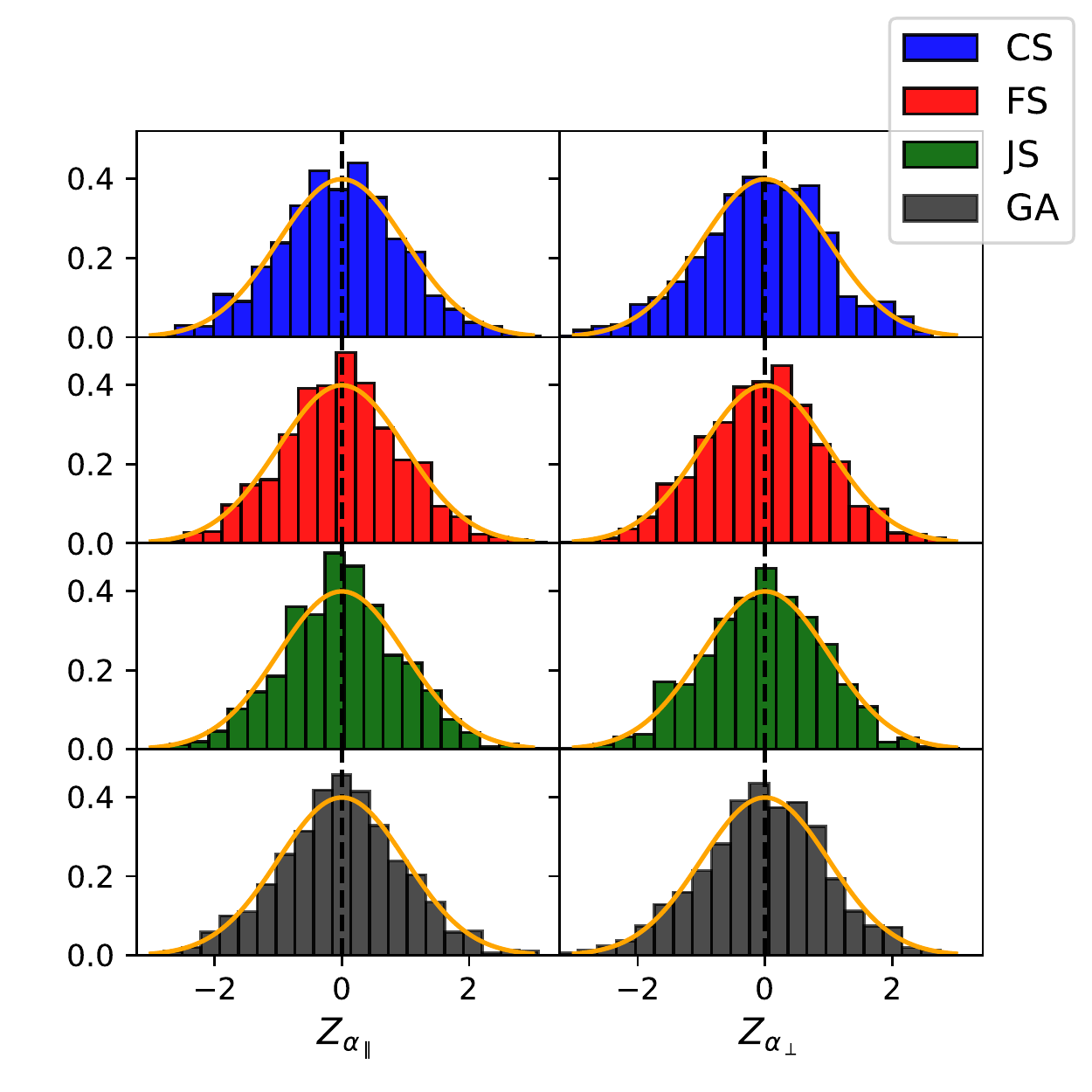}
	\caption{Distributions of the pull $Z_i = \left(x_i - \langle x_i\rangle\right)/\sigma_i$ for the parameters $\left(\alpha_\parallel,\alpha_\perp\right)$ obtained obtained from the CS, FS, GA and JS analyses of the 1000 EZmocks sample. Orange curves describing a Gaussian distribution with zero mean and the standard deviation of the corresponding pull are shown for comparison. The dotted lines are centered on zero.}
	\label{fig:pull_asym}
\end{figure}

Figure \ref{fig:pull_asym} displays the distributions of the pull $Z_i$
for $\apara$ and $\aperp$. 
For all four methods, the mean of the pull $\langle Z_i \rangle$ 
is centered in zero to a few percent level, suggesting again that there are 
no systematic bias in the estimated $\alpha$.
Moreover the standard deviation of the pull $\sigma\left(Z_i\right)$ is
smaller than one for both parameters in all analyses indicating a 
slight overestimation of the errors.
This overestimation is more important for the parameter $\alpha_\parallel$, 
and in JS in general. As the standard deviation of the pulls are only 
a few percent away from unity, we do not attempt to correct these effects. 
If any overestimation of uncertainties is real, not correcting for it can be 
considered as an conservative approach. 
Yet, we see that for the parameter $\alpha_\parallel$ the overestimation 
of the errors reaches 10\% in JS. While this seems to be significant, 
this result can be a consequence of the Gaussian assumption for the 
distribution of the parameters. 

To investigate the Gaussian nature of the distribution of 
the $\alpha$'s parameters, we perform a D’Agostino-Pearson’s test
\citep{dagostino_omnibus1971, dagostino_tests_nodate} on the sample 
of best-fit parameters using the \textsc{Scipy.stats} 
library\footnote{\url{https://docs.scipy.org/doc/scipy/reference/stats.html}}. 
The test combines the high-order statistical moments skewness 
and kurtosis to test the hypothesis that the two parameters are 
normally distributed. The statistical quantity $K_2$ is constructed 
as a combination of the renormalized skewness and kurtosis such 
that $K_2$ asymptotically follows a $\chi ^2$ law with two degrees 
of freedom. When the skewness and kurtosis simultaneously deviate 
from 0 (Fisher kurtosis) $K_2$ gets larger. The p-value for the
$K_2$ statistic is then to be compared with the risk $a$ usually 
set to 5$\%$ for a Pearson test. As the $a$ is the risk to reject 
the null hypothesis $H_0$ while it is true, the p-value should be
less than $a$ to reject $H_0$. Here the null hypothesis is that 
the $\alpha$'s sample comes from a normal distribution. 
Table~\ref{tab:DagonistoPearson_test} shows the results of the Pearson tests on the $\alpha_\perp$, and $\alpha_\parallel$ distributions. 
Independently of the method, the Gaussian distribution hypothesis can be rejected for the parameter $\alpha_\parallel$ only, for which $p/a < 1$.
This result indicates that for $\alpha_\parallel$, the smaller values of $\sigma(Z_i)$ observed in Table~\ref{tab:stats} might be partially induced by the non-Gaussianity of the $\alpha$ distributions and not simply by an overestimation of the uncertainties.

\begin{table}
    \centering
    \caption{Results of D’Agostino and Pearson’s test of normality over the 
    $\alpha$'s distributions. $K_2=z_k^2 + z_s^2$ is a combination of the 
    skewness and kurtosis normalized coefficients. The p value of the test 
    is a 2-sided chi squared probability for the Gaussianity hypothesis test. 
    The parameter $a$ that we set at 5\% is the risk (usually named $\alpha$ 
    in the literature) that we reject the hypothesis $H_0$ of normality 
    while it is true. }
    \begin{tabular}{ccccc}
    \hline
    \hline
    &\multicolumn{2}{c}{$\alpha_{\perp}$} & 
     \multicolumn{2}{c}{$\alpha_{\parallel}$} \\
    Analysis & $K_2$ & $p/a$ & $K_2$& $p/a$ \\  
    \hline

    CS &5.55 & 1.249 &17.89 & 0.003 \\
    FS &2.92 & 4.65 &16.92 & 0.004 \\
   GA &3.42 & 3.623 &11.14 & 0.076 \\
    JS &2.64 & 5.33 &11.91 & 0.052 \\

    \hline
    \hline
    \end{tabular}
    \label{tab:DagonistoPearson_test}
\end{table}

\begin{figure*}
  \centering
  \includegraphics[width=0.4\textwidth]{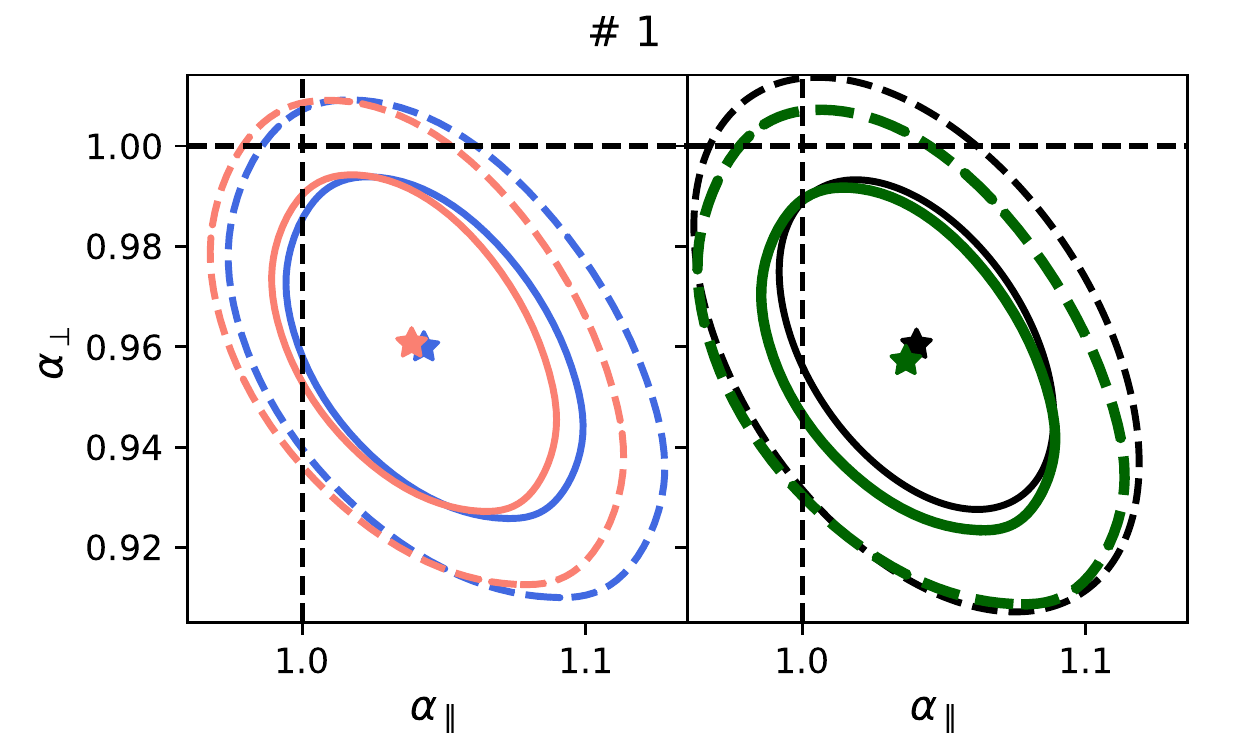}  
  \includegraphics[width=0.4\textwidth]{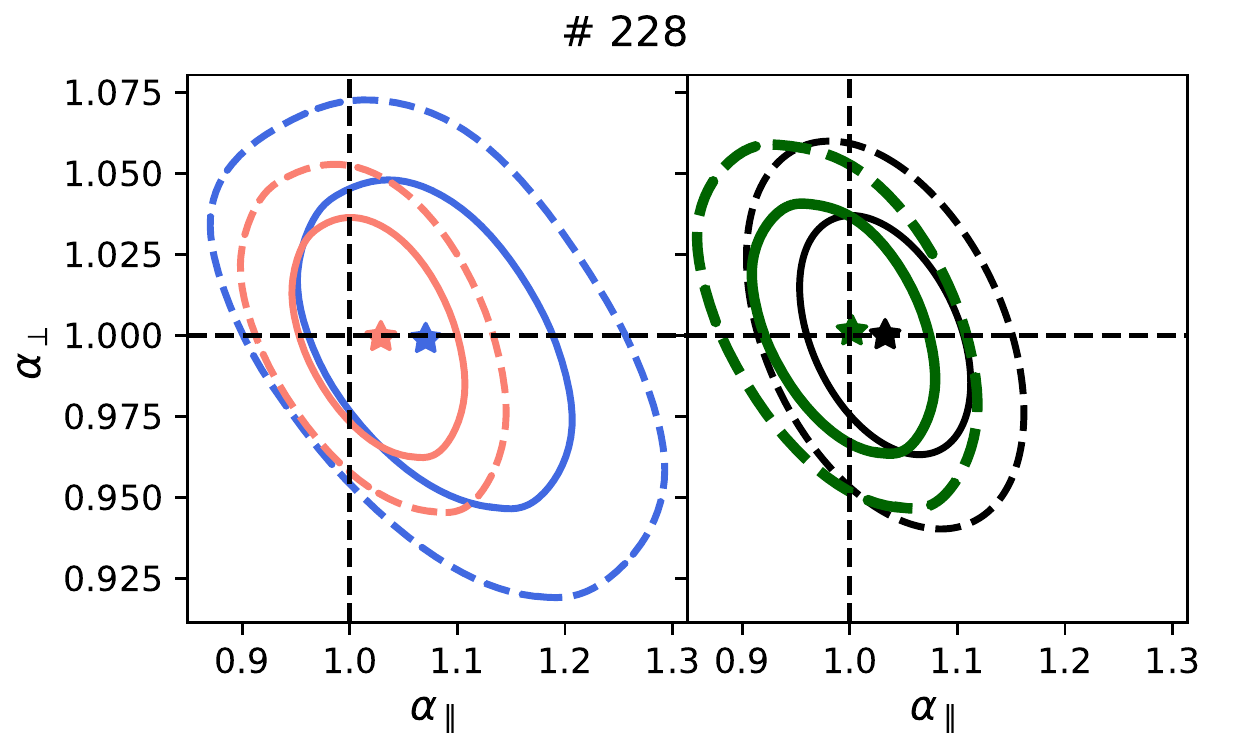}  
  \includegraphics[width=0.4\textwidth]{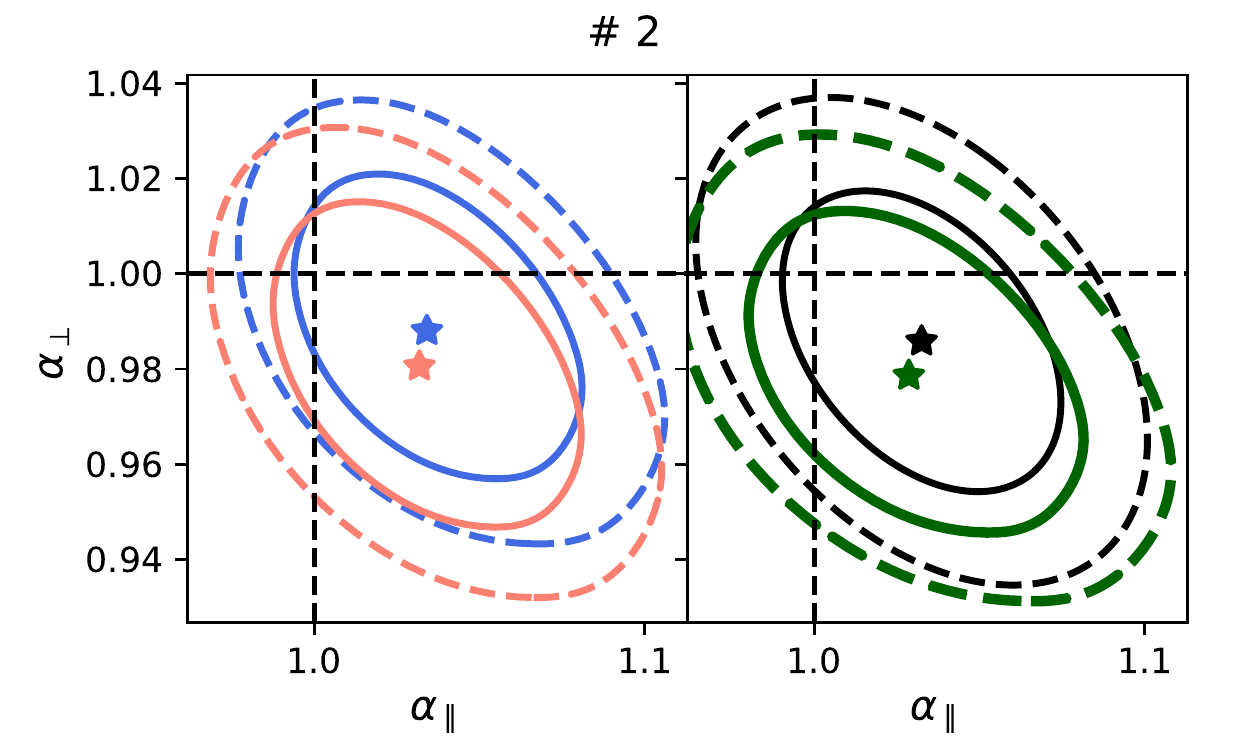}  
  \includegraphics[width=0.4\textwidth]{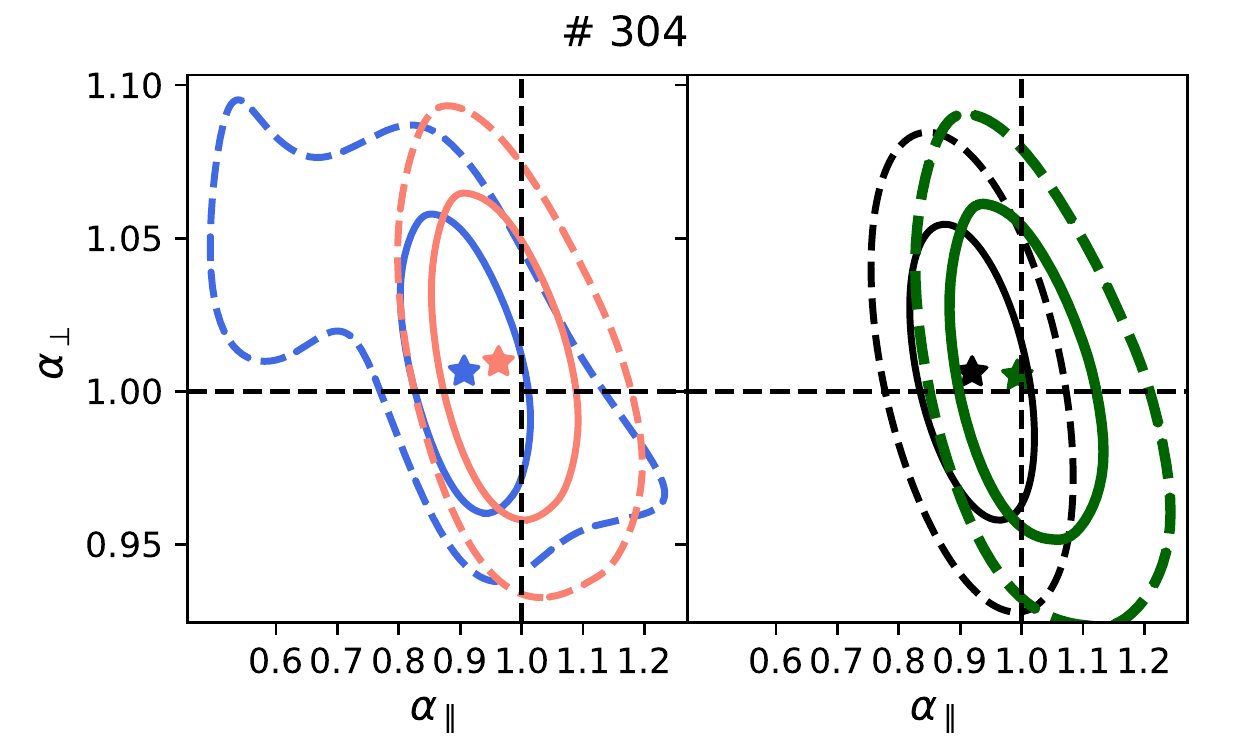}  
  \includegraphics[width=0.4\textwidth]{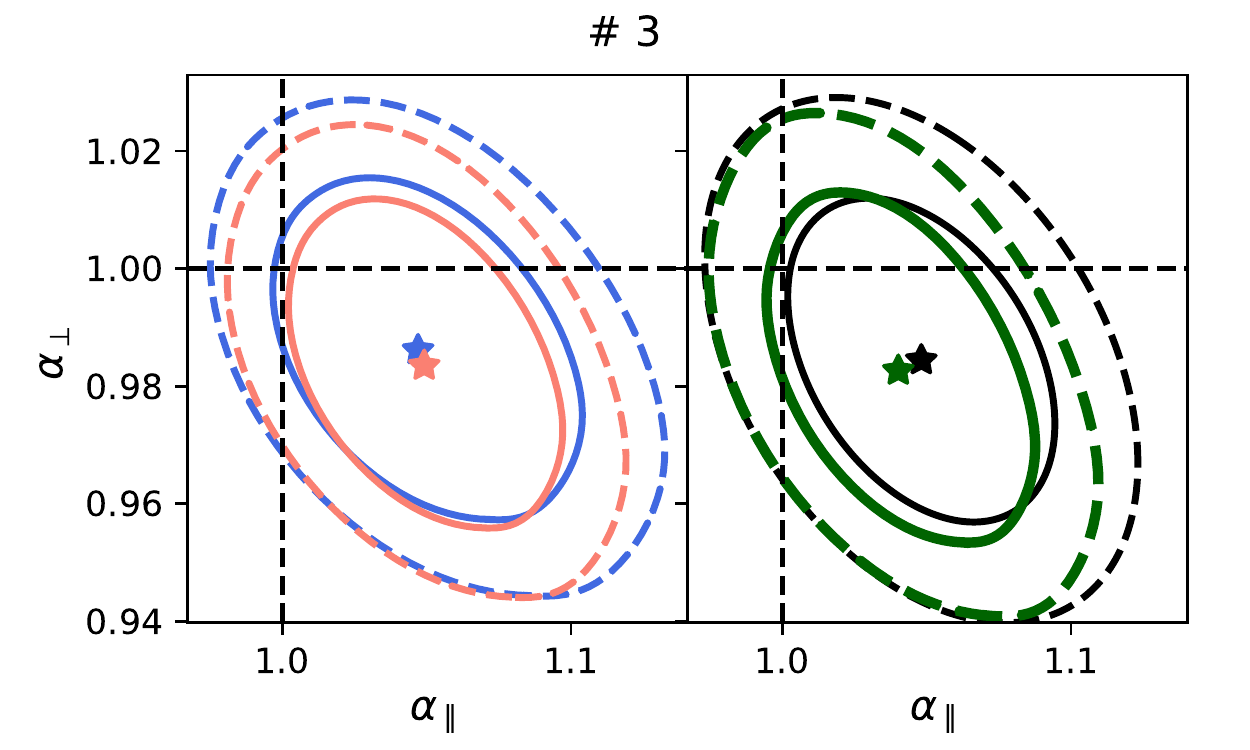}  
  \includegraphics[width=0.4\textwidth]{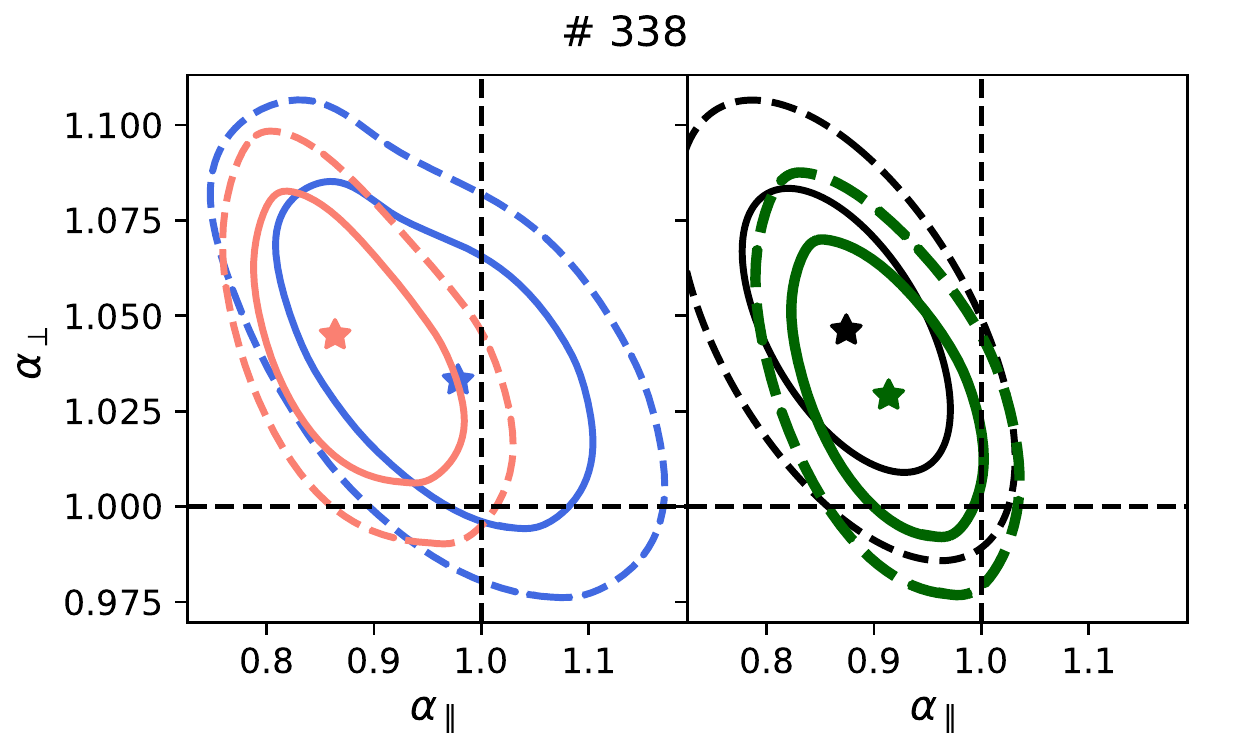}  
  \includegraphics[width=0.4\textwidth]{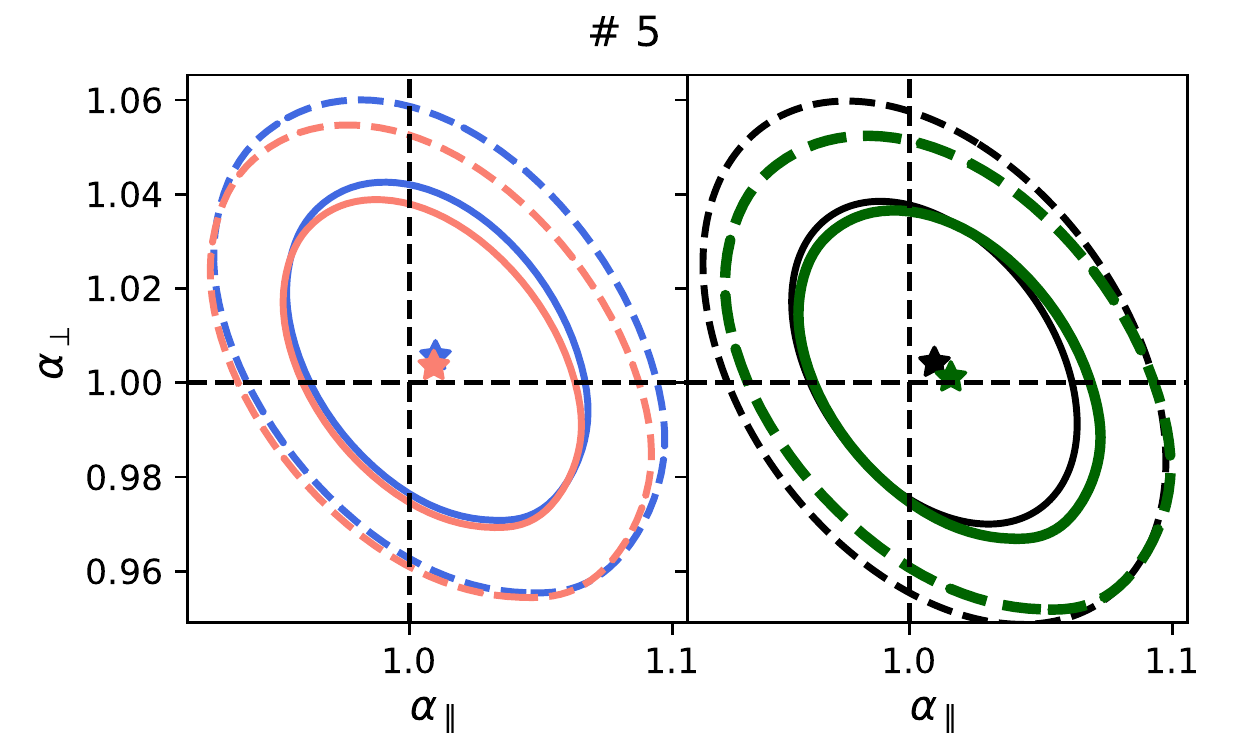}  
  \includegraphics[width=0.4\textwidth]{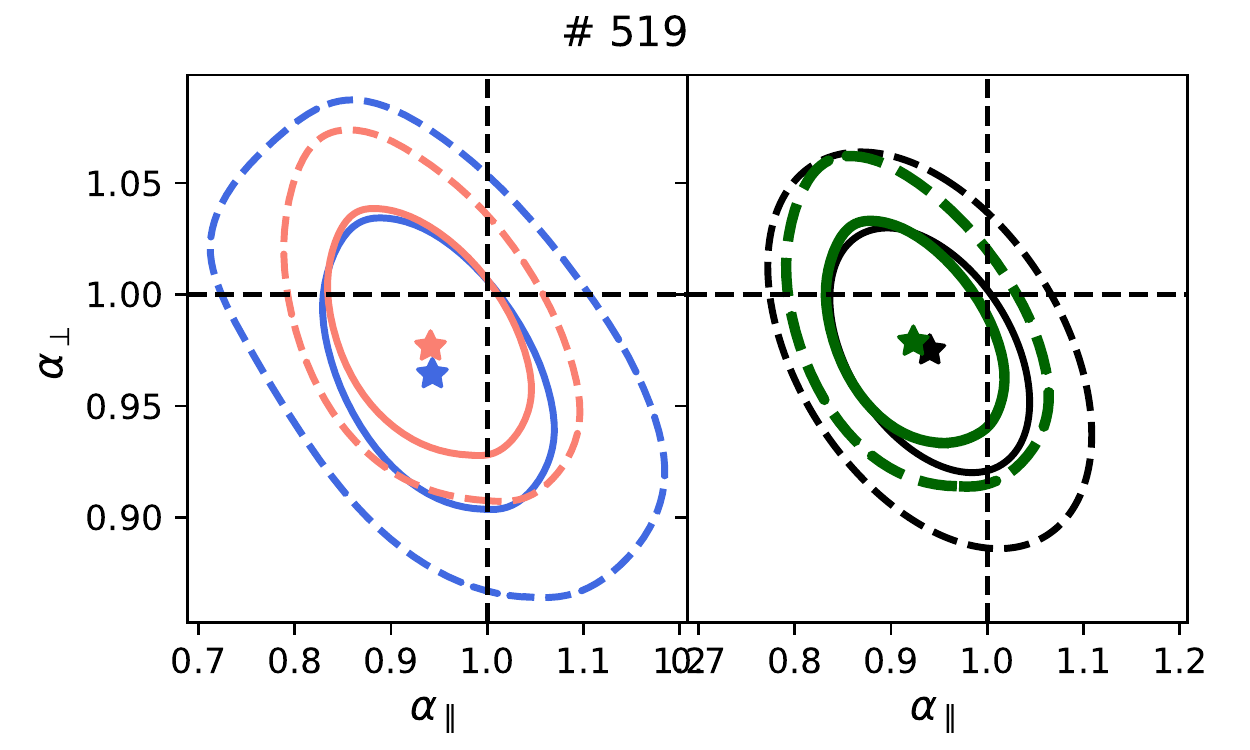}  
\caption{Comparison between the different methods, using some EZmock realisations (mock tag above each plot), when the contours found in configuration or Fourier space are Gaussian(left) and non Gaussian(right). Red contours are for FS results, blue for CS, green for JS and black for GA. The expected value is the intersection of the dotted black lines, and the best fit values are described by a star for the JS and GA methods. We can see how the JS method yields better combined results that are not necessarily Gaussian.}
\label{fig:non_gaussian_contours}
\end{figure*}

Figure~\ref{fig:non_gaussian_contours} shows two dimensional confidence levels 
for a few realisations of mocks. Red contours are for FS results, blue for CS, green for JS and black for GA. The left panels are cases where the likelihoods can be approximated by a 2D Gaussian distribution while the right panels the opposite. 
In these examples, we can see how the JS fits are a better description for the non-Gaussian cases, where the final consensus are not necessarily Gaussian.

\section{Application to eBOSS DR16 data}
\label{sec:results_data} 

In this section, we apply our methodology to the eBOSS LRG sample, 
described in section~\ref{sec:data:eboss}, letting free only $\alpha_\parallel$, $\alpha_\perp$, the linear density bias $b$ and the broadband parameters. 

\begin{figure*}
	\centering
	\includegraphics[width=0.9\textwidth]{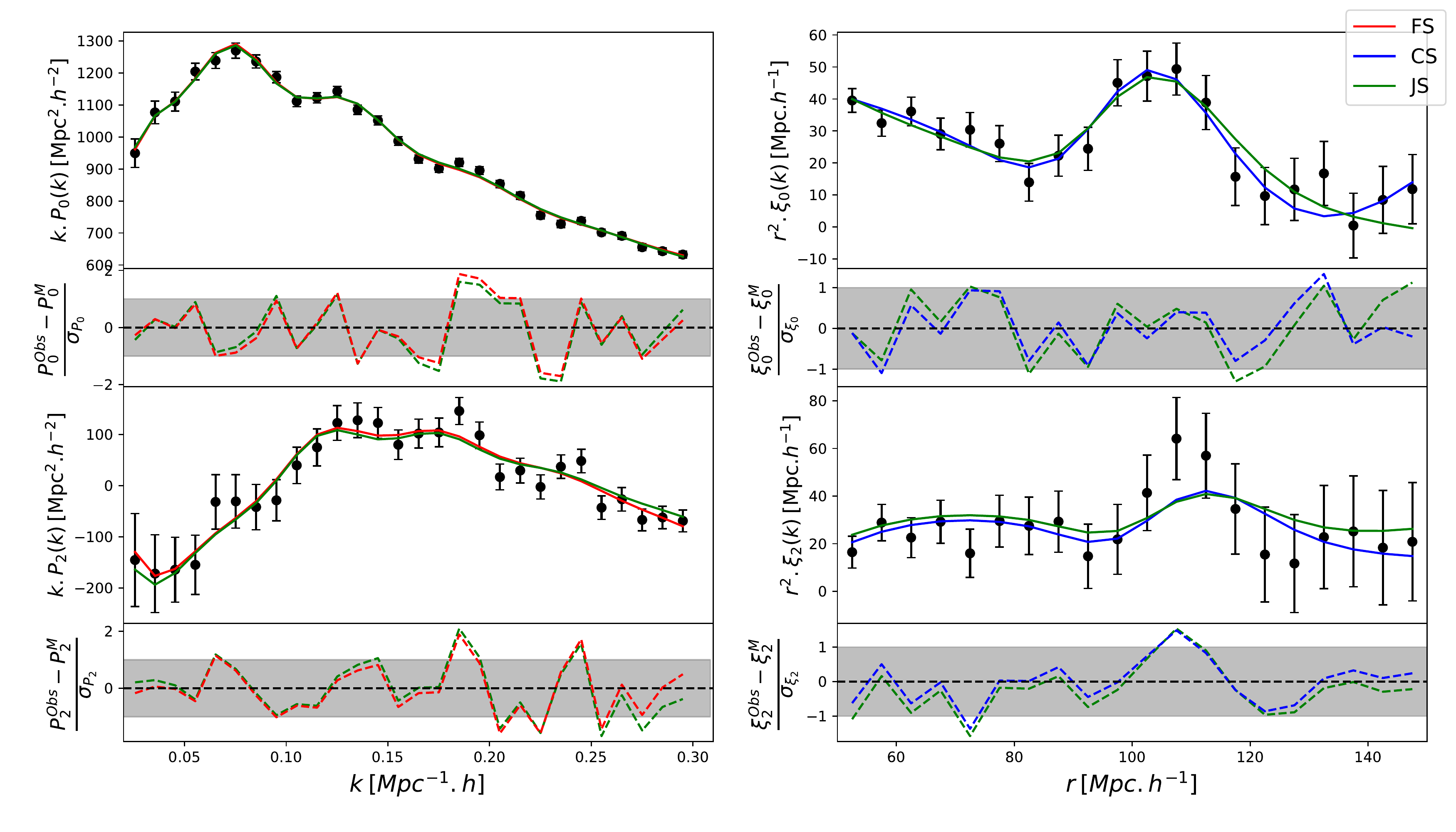}
	\caption{Best monopole and quadrupole models for the CS, FS and JS analysis. The residues with respect to the measurements are standardized for each point by the errorbar. The grey shaded area correspond to a 1$\sigma$ difference.}
	\label{fig:data_multipole_fit}
\end{figure*}

Figure~\ref{fig:data_multipole_fit} shows the best-fit 
models for the power spectrum and correlation function 
multipoles, for three analyses: FS, CS and JS (note that 
we do not show a GA best-fit model as GA results are derived using the CS and FS best-fit). 
The residuals between in the bottom panels show excellent agreement between the JS model and
the individual models for FS and CS, particularly 
over the BAO features. Furthermore, as the global amplitudes 
driven by the parameter $b$ are sensibly the same, the small 
differences between the models arise mainly from the
different broadbands. This is specially true for the 
correlation function as it is composed of fewer data bins.

\begin{figure}
	\centering
	\includegraphics[width=1.\columnwidth]{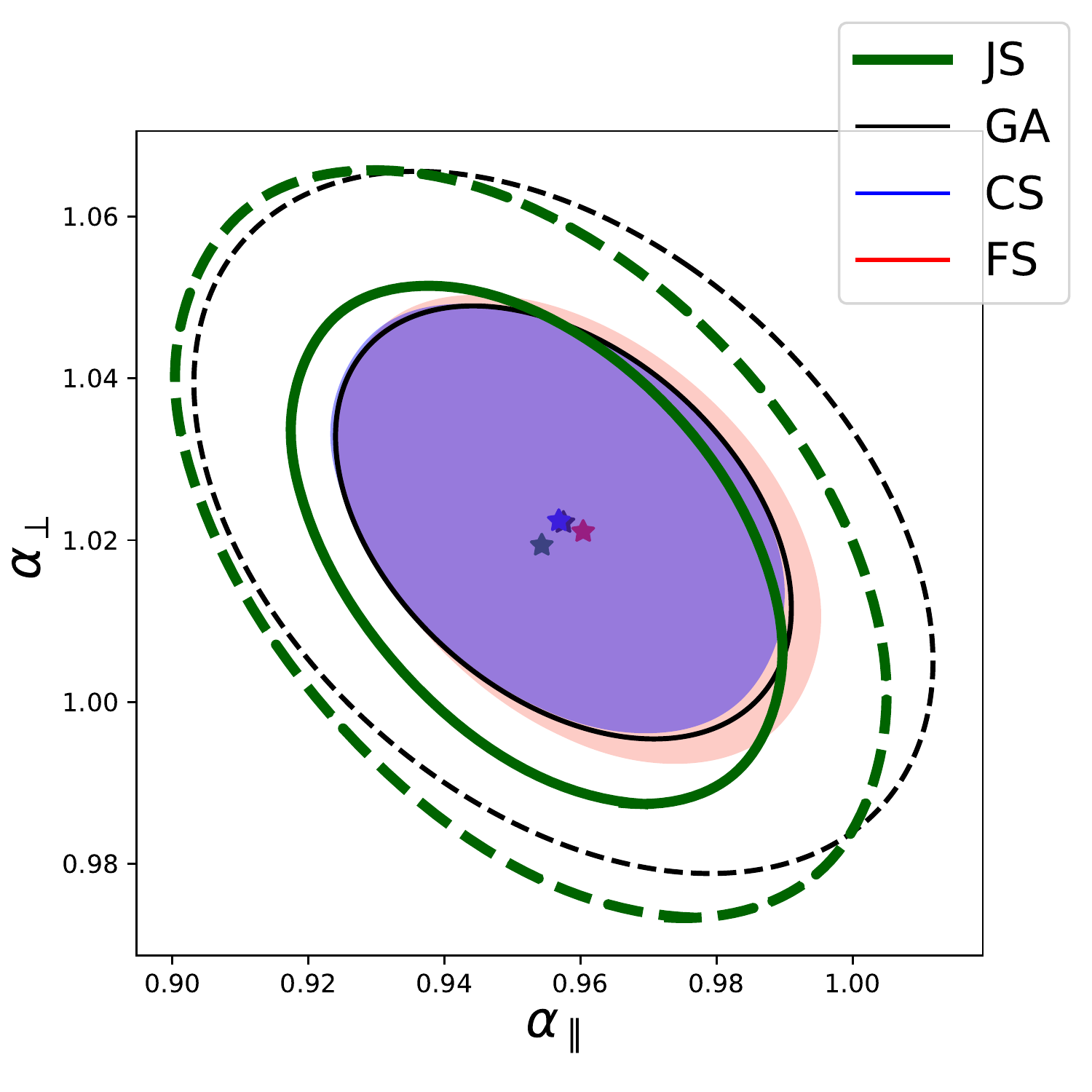}
	\caption{Confidence contours of the parameters posterior distributions for the four analysis. For the CS and JS, the 1$\sigma$ contours are filled in blue and red respectively. For the JS and GA we give the 1$\sigma$ and 2$\sigma$ contours for comparison in green and black. The best values are designated with a star. }
	\label{fig:contours_data}
\end{figure}

Figure~\ref{fig:contours_data} shows the two dimensional 
68 and 95 percent confidence levels for CS, FS, GA and JS. 
Once again we see an excellent agreement between the four analysis, 
with the JS giving slightly looser constraints. 
We also notice small differences in the inclination of the contours. 
The correlation coefficients for the CS, FS, GA and JS are 
respectively (-0.395, -0.404, -0.396, -0.445). Those values are 
consistent with the correlations found with 1000 EZmock realisations: 
(-0.380, -0.395, -0.383, -0.420). Note that this coefficient for the GA is 
highly dependent on the quality of the estimated correlation matrix of the 
parameters estimated from the 
EZmock sample and shown in Figure~\ref{fig:alpha_corr_tot}. 
We find a higher correlation 
coefficient using the JS analysis, resulting in a thinner and steeper contours. 
This have for effect to increase the errors on the individual parameters 
without varying much the Figure-of-Merit (area of the contour, 
see Appendix~\ref{app:nmock} for a discussion on the FoM).

\begin{table*}
    \caption{Best-fit BAO parameters from the eBOSS DR16 LRG sample for different methods: configuration-space (CS), Fourier-space (FS), the Gaussian combination (GC) and our new joint-space fit (JS). The $\alpha$ values are relative to our fiducial cosmology for which $D_M/r_d(z_{\rm eff} = 0.7) = 17.436$ and $D_H/r_d(z_{\rm eff} = 0.7) = 20.194$. }
    \centering
        \begin{tabular}{ccccccc}
            \hline
            \hline
             Analysis & $\aperp$ &  $\apara$ & $\chi^2_{\rm min}/n_{\rm dof}={}_{r}\chi^2_{\rm min}$ & $p_{\rm value}$ & $D_H/r_d$ & $D_M/r_d$ \\
              \hline
            CS & 1.022 $\pm$ 0.018 & 0.957 $\pm$ 0.022 & $36.8/(40-11)=
1.27$ & 0.151 & 19.32 $\pm$ 0.44 &  17.83 $\pm$ 0.30\\
            FS & 1.021 $\pm$ 0.019 & 0.960 $\pm$ 0.024 & $57.6/(56-11)=1.28$ & 0.098 & 19.39 $\pm$ 0.47 &  17.80 $\pm$ 0.33\\
            GA & 1.022 $\pm$ 0.018 & 0.957 $\pm$ 0.022 & -     & -     & 19.33 $\pm$ 0.45 &  17.82 $\pm$ 0.31\\
            JS & 1.019 $\pm$ 0.022 & 0.954 $\pm$ 0.024 & $97.7/(96-19)=
1.21$ & 0.056 & 19.27 $\pm$ 0.48  & 17.77 $\pm$ 0.37\\
            \hline
            \hline
        \end{tabular}
        \label{tab:results_data}

\end{table*}

Table~\ref{tab:results_data} summarises the results of the four 
different analysis. The reduced $\chi^2$ values are 1.28, 1.27 and 
1.21 respectively for FS, CS and JS indicating relatively good 
adjustment to the data without over fitting from the broadband. 
In each case, the $p_{\rm value}$ indicates a valid fit (if compared 
with a risk of 5$\%$). The best fit values for the parameters
$(\alpha_\perp,\alpha_\parallel)$ are all in agreement according 
to their respective 1$\sigma$ error bar. For both parameters, 
the JS analysis gives the larger uncertainty. Rescaling with 
the fiducial cosmology, we also derive the corresponding physical 
quantities  $D_M/r_d$ and $D_H/r_d$. Note that our best fit results 
do not exactly match the BAO constraints obtained in 
\cite{BautistaCompletedSDSSIVExtended2021} and 
\cite{GilMarinCompletedSDSSIVExtended2021}, 
neither in configuration space nor in Fourier space, 
where the GA consensus gives $D_H/r_d = 19.33 \pm 0.53$ and $D_M/r_d =17.86 \pm 0.33$
(third entry of Table 14 of \citealt{BautistaCompletedSDSSIVExtended2021}). 
Those discrepancies arise from the slightly different non linear 
broadening parameters ($\Sigma_\parallel, \Sigma_\perp$) 
and a more flexible broadband polynomial expansion 
$(i_{\rm min}, i_{\rm max}) = (-2, 1)$ to be compared to $(-2, 0)$
for \citet{BautistaCompletedSDSSIVExtended2021} and $(-1, 1)$ for 
\citet{GilMarinCompletedSDSSIVExtended2021}.

\section{Conclusions}

This work introduced a new BAO analysis of the DR16
eBOSS LRG sample, using Fourier and configuration-space information
simultaneously. 
We compared the joint space (JS) analysis with the commonly used 
Gaussian approximation (GA) method, which combines results from the two 
spaces (Fourier and configuration) at the parameter level. 
The main advantage of the JS method is that it does not require 
any Gaussian assumption for the likelihood profiles. 
While the GA method is accurate only if the individual likelihoods 
to be combined are both Gaussian, yielding only Gaussian posterior
distributions. 

We assessed the systematic biases and errors of both methods by 
applying JS and GA to a set of 1000 EZMocks, which reproduce 
the eBOSS DR16 LRG sample properties. 
Compared to GA, JS provides a more accurate estimation of the 
acoustic scale by lowering the systematic shift of $\alpha_\parallel$ 
with respect to its expected value. 
while GA has by construction a better 
precision (smaller error bars). 
Moreover, the JS offers a better 
control over variations of the analysis. 
Indeed the same mocks as the 
ones used to estimate the covariance are commonly used to perform 
systematic studies. Because of this, the standard deviation of any 
parameter should be rescaled with the correction factor $m_2$ 
(see Table \ref{tab:correction_factor}), which is not properly 
defined for the GA. 
One should note that because of larger size of the data vector, 
the JS methods requires a sufficiently large mock sample to 
estimate the covariance. 
However we found that the constraints are quite stable, 
and the number of mocks needed is not much larger than for 
a regular configuration or Fourier analysis.

We applied the different analysis to the eBOSS LRG sample 
and found consistent results (see Table \ref{tab:results_data}). 
As expected from the statistical study the JS gives slightly 
looser constraints and a larger correlation coefficient. 
Despite providing looser constraints on cosmological parameters 
than the standard GA, we believe that JS is a more robust and 
reliable method for modelling the clustering signal by correctly 
accounting for the correlation between configuration and Fourier 
space and not relying on the Gaussianity of the parameter likelihoods.

One specific feature of the JS in a BAO analysis is that the 
broadband terms are independent between spaces. Hence, the JS 
introduces more nuisance parameters to be fitted simultaneously. 

We are currently working on extending joint-space fits for measurements of 
the growth-rate of structures using information from the full shape of the power spectrum and correlation function. This work can also be extended
to joint fits between pre and postreconstruction catalogues, 
as already performed in Fourier space by \cite{gil-marinHowOptimallyCombine2022}.

\begin{acknowledgements}
We would like to thank Hector Gil-Marín, Ashley Ross,  Elena Sarpa, for useful 
discussions. 
The project leading to this publication has received funding from 
Excellence Initiative of Aix-Marseille University - A*MIDEX, 
a French ``Investissements d'Avenir'' program (AMX-20-CE-02 - DARKUNI).

\end{acknowledgements}

%
%

\bibliographystyle{aa}
\bibliography{Biblio}

\appendix

\section{Fit with hexadecapole}
\label{sec:annex:hexa}

Table \ref{tab:hexadec} shows the results for the fit of the averaged 1000 EZmocks, with and without using the hexadecapole. For every analysis the use of the additional information results in a slight increase of the systematic bias and of the errors bars. The looser constraints mainly arise from the larger size of the data vector when using hexadecapole. Indeed a larger amount of data bins results in an increase of the Whishart bound through the rescaling parameter $m_1$ (especially for the JS analysis). Thus we decide not to use the hexadecapole information in this work.

\begin{table}
\addtolength{\tabcolsep}{-5pt}    
    \centering
    \caption{Results for the fit of the 1000 EZmocks stack with and without using the hexadecapole information. For each parameter, we show the bias $\Delta_\alpha \equiv (\alpha_i - \alpha_{\rm exp})$ and the estimated uncertainties $\sigma_\alpha$.}
    \begin{tabular}{c|cccc|cccc}
    \hline
    \hline
    &\multicolumn{4}{c}{$\ell = (0,2)$} \vline&\multicolumn{4}{c}{$\ell = (0,2,4)$}\\
    \hline
    &\multicolumn{2}{c}{$\alpha_{\perp}$} & 
     \multicolumn{2}{c}{$\alpha_{\parallel}$}\vline&
     \multicolumn{2}{c}{$\alpha_{\perp}$} & 
     \multicolumn{2}{c}{$\alpha_{\parallel}$} \\
    Analysis & $\Delta_\alpha $ & $\sigma_\alpha$ & $\Delta_\alpha $ & $\sigma_\alpha$& $\Delta_\alpha $ & $\sigma_\alpha$& $\Delta_\alpha $ & $\sigma_\alpha$ \\ 
    & [$10^{-2}$] & [$10^{-2}$]  & [$10^{-2}$] & [$10^{-2}$] & [$10^{-2}$]& [$10^{-2}$]& [$10^{-2}$]& [$10^{-2}$] \\

    \hline

    CS&0.17&2.28&0.24&3.5&0.18&2.3&0.25&3.54 \\
    FS&0.09&2.25&0.39&3.42&0.11&2.27&0.39&3.44 \\
    GA&0.13&2.23&0.32&3.35&0.14&2.24&0.33&3.36\\
    JS&0.09&2.34&0.13&3.49&0.13&2.4&0.13&3.6 \\

    \hline
    \hline
    \end{tabular}
    \label{tab:hexadec}
\end{table}

\section{Prereconstruction results}
\label{sec:annex:prerecon}

\begin{figure}
\centering
\includegraphics[width=0.8\columnwidth]{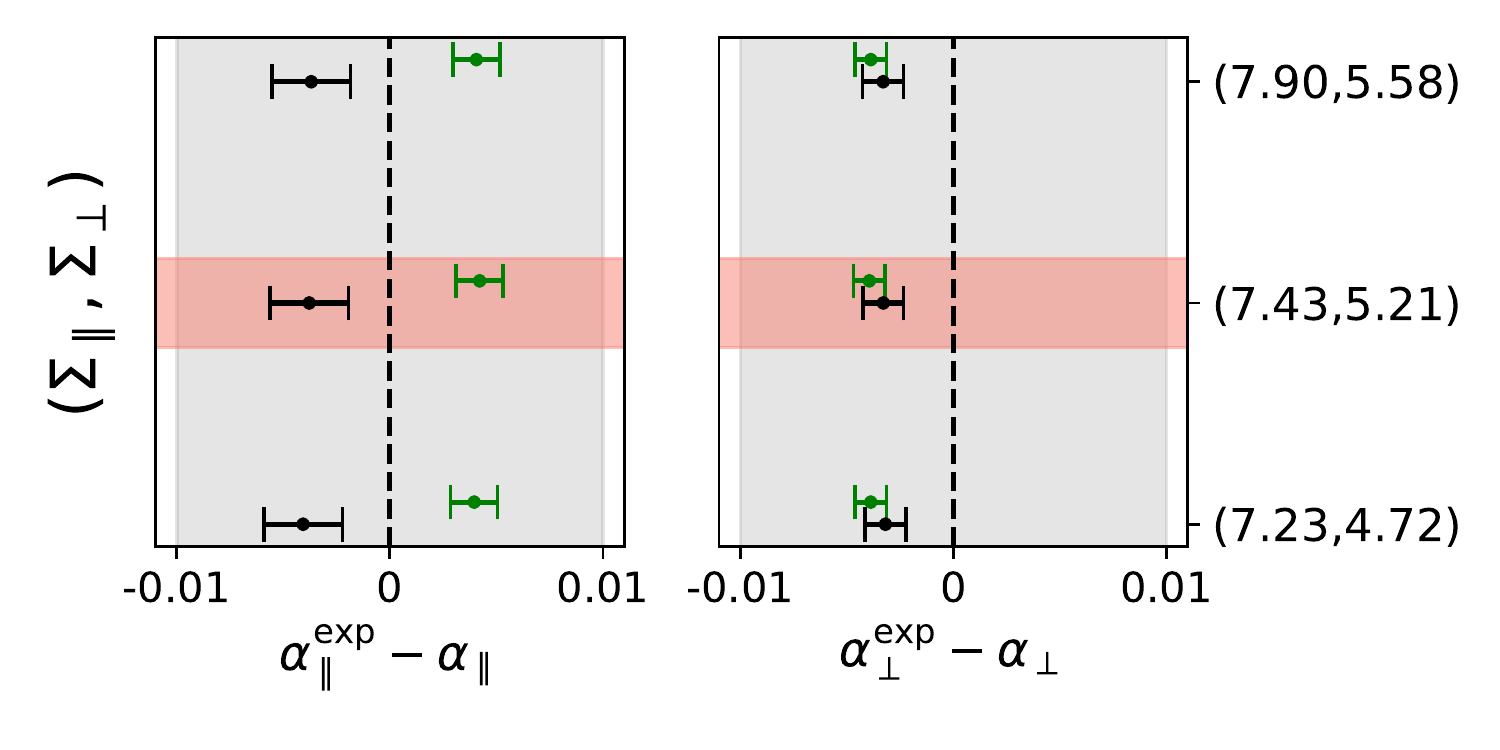} 
\includegraphics[width=0.8\columnwidth]{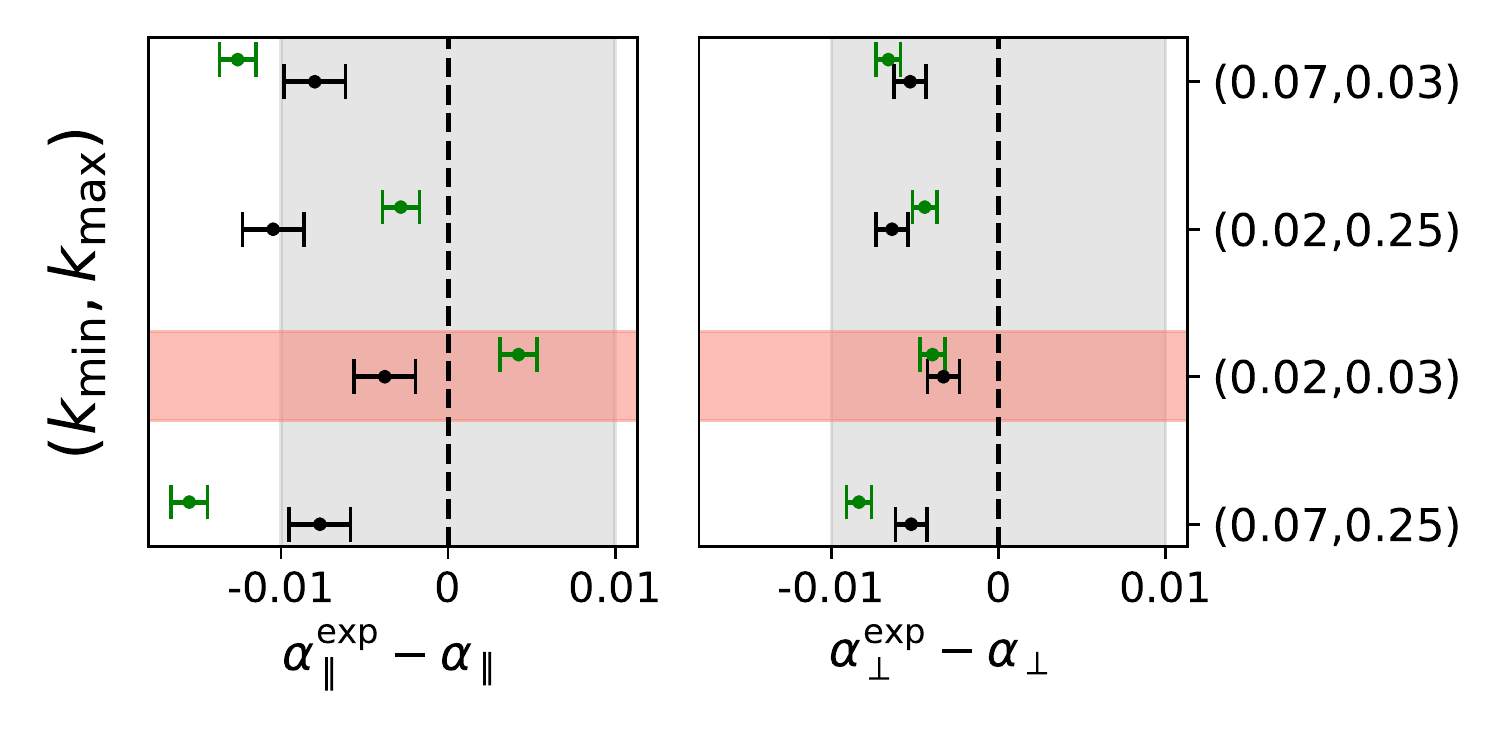}  
\includegraphics[width=0.8\columnwidth]{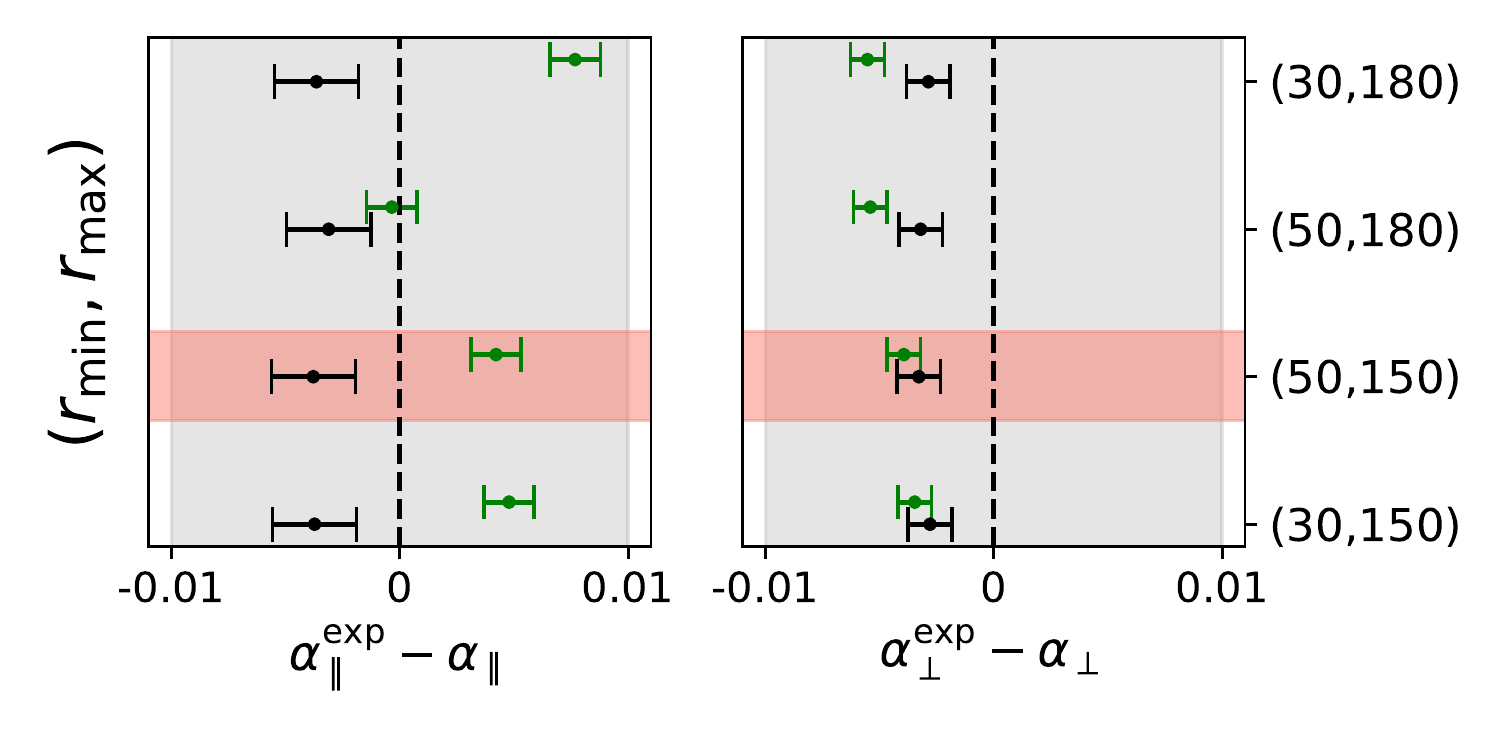}  
\includegraphics[width=0.8\columnwidth]{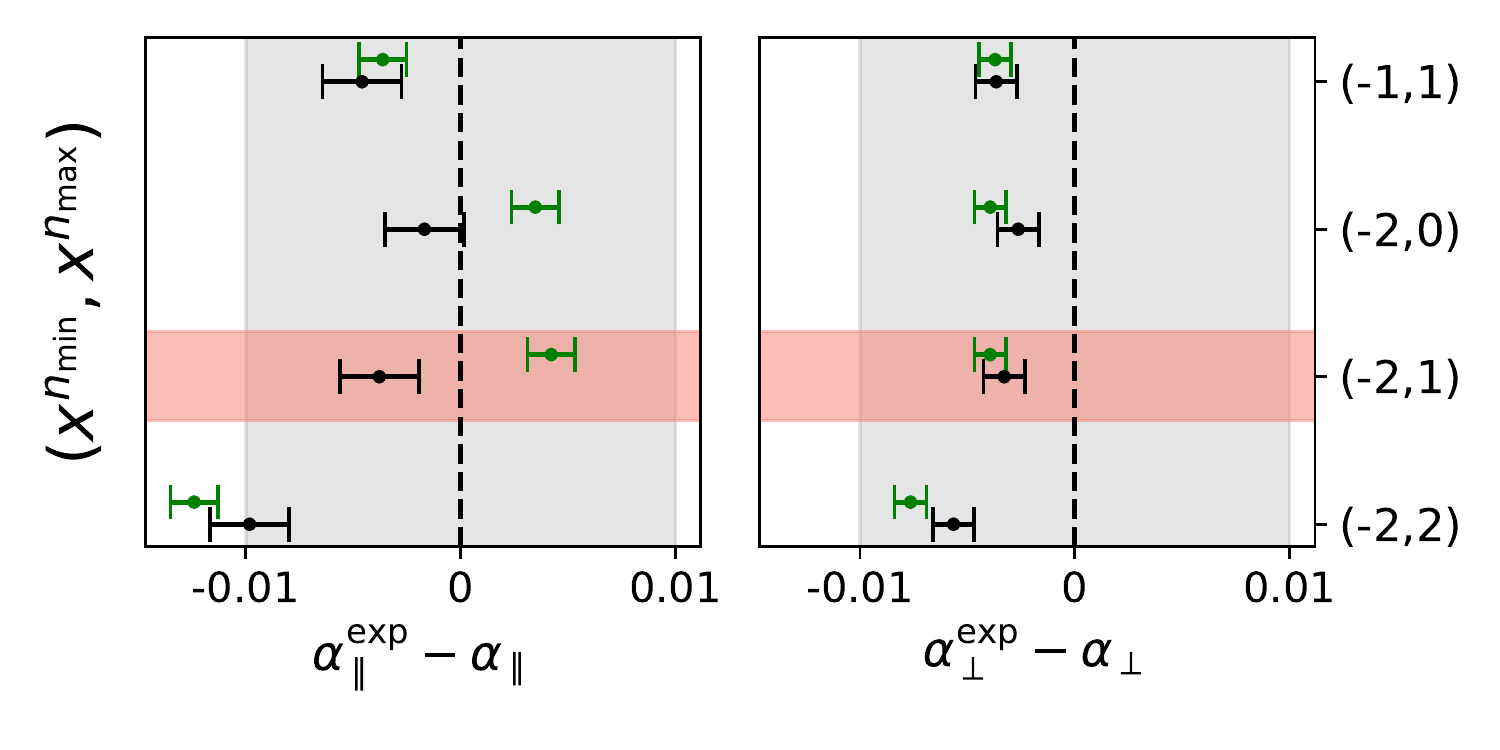}  
\includegraphics[width=0.8\columnwidth]{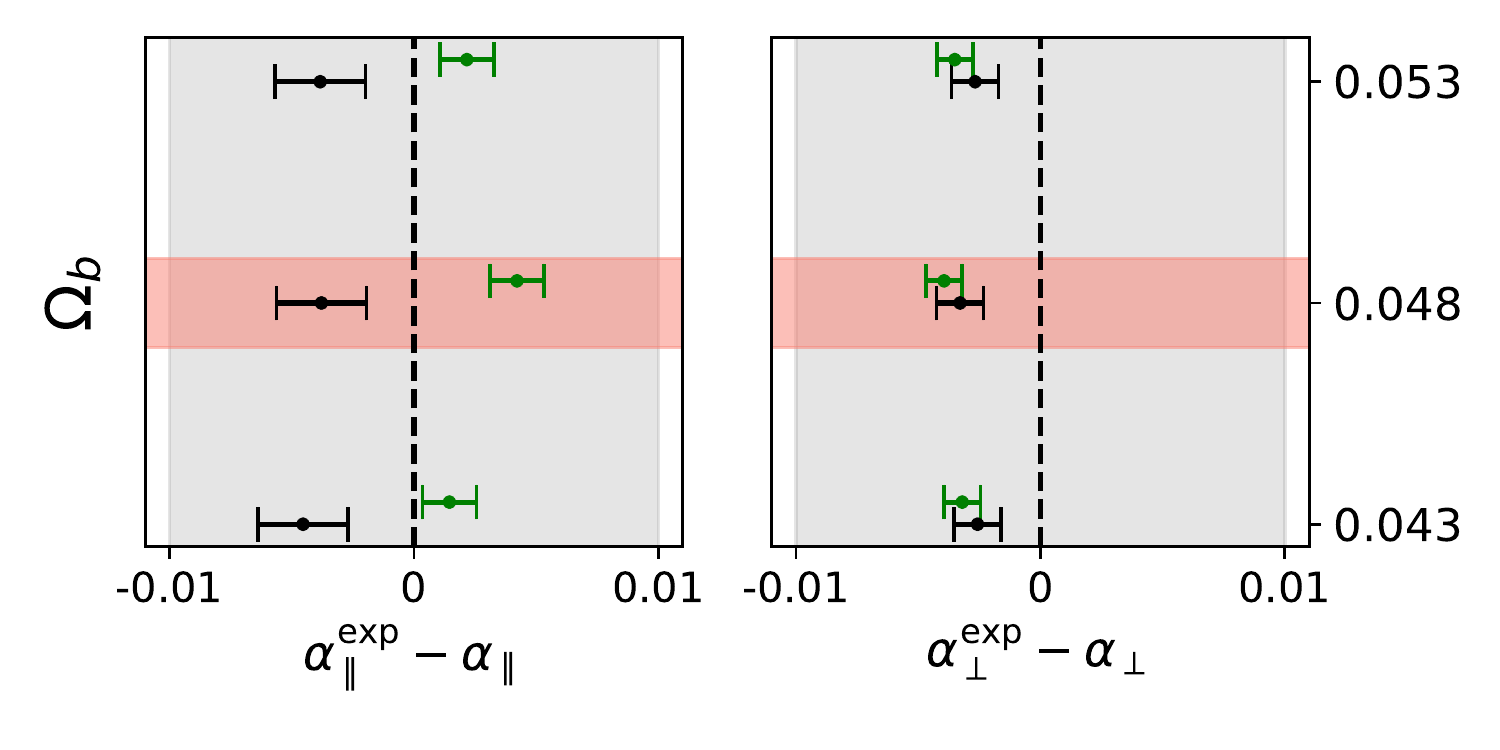}  
\includegraphics[width=0.8\columnwidth]{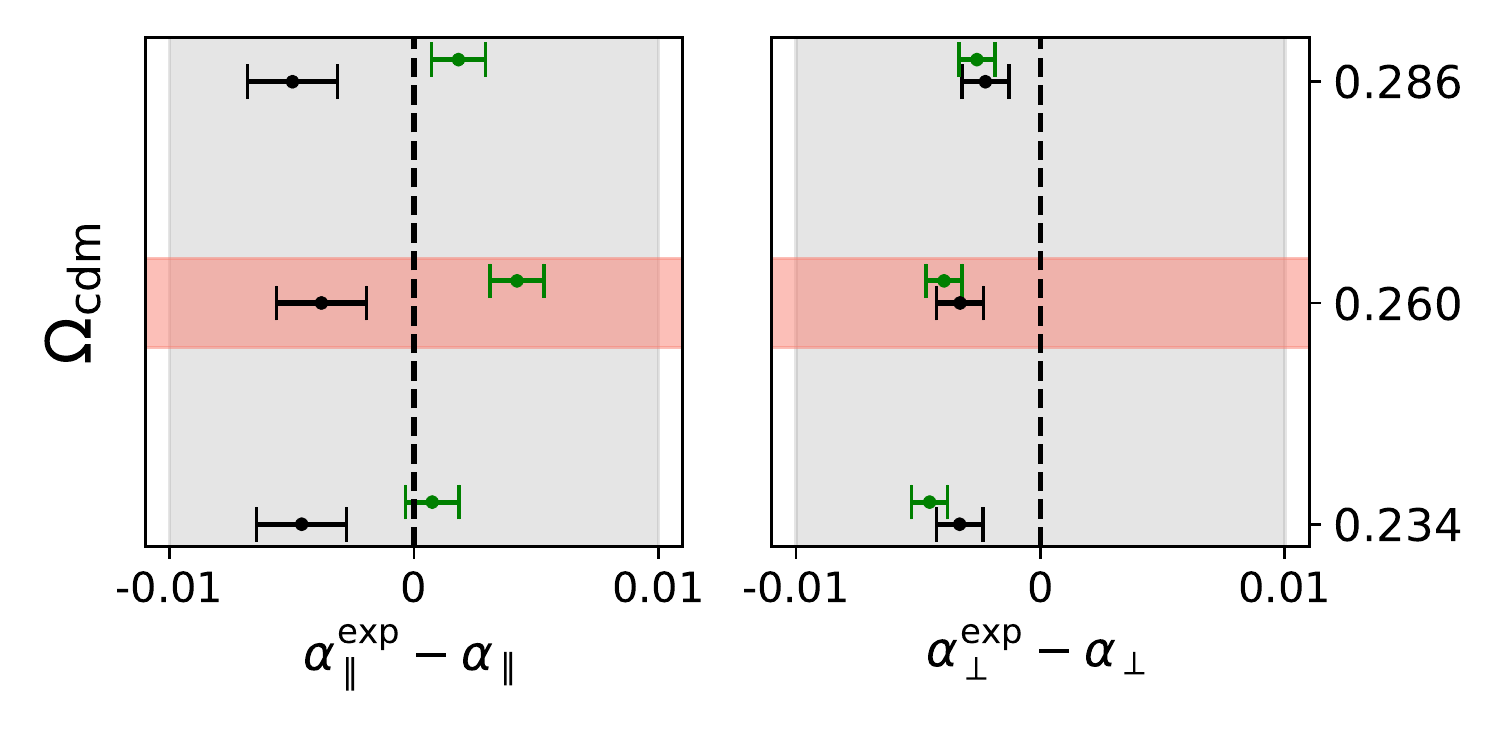}  
\caption{Impact of the choice of fitting scales, fiducial cosmology, broadening $\Sigma$'s parameters and polynomial broadband order in the recovered values of the parameters $\apara$ and $\aperp$. Each point is the best-fit from the average of the 1000 prereconstruction EZmocks. The JS method is in green and GA method is in black. The grey shaded areas correspond to a 1 percent error and red shared areas indicate the fiducial choices of our analysis.} 
\label{fig:syst_prerecon}
\end{figure}

We perform fits on average correlations and individual mocks for the prereconstruction sample. Figure~\ref{fig:syst_prerecon} presents the impact of different parameters choice on the best fit values of $\alpha_\parallel$ and $\alpha_\perp$ for the GA (in black) and JS (in green) analysis. In every plot, a grey shaded area represents the 1 percent deviation from the expected value. Both analysis appears to be more dependant on the range of scales used (particularly in fourier space) as the small scales non linearities are not well modeled. However, without the reconstruction procedure that requires a fiducial cosmology, the dependence on $\Omega_b$ and $\Omega_{\rm cdm}$ is reduced. The two analysis gives different results for the parameter $\alpha_\parallel$, the JS analysis tends to give a positive bias while the GA gives a negative.

\begin{figure}
	\centering
	\includegraphics[width=1\columnwidth]{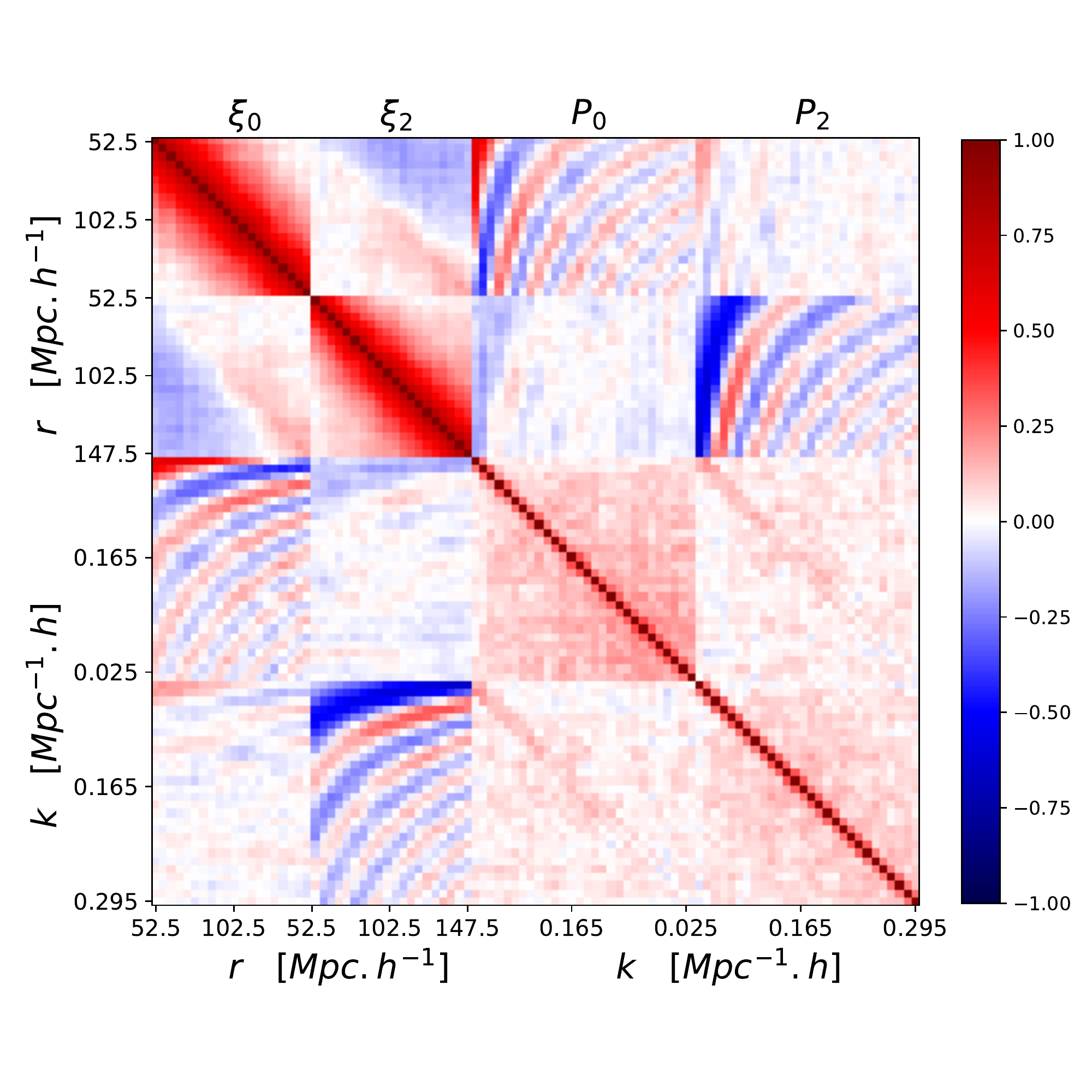}
	\caption{Correlation matrix of the correlation function $\xi_\ell$ and power spectrum multipoles $P_\ell$ estimated using the 1000 independent measurements of prereconstructed EZmocks.}
	\label{fig:covariance_pre}
\end{figure}

\begin{figure}
	\centering
	\includegraphics[width=0.8\columnwidth]{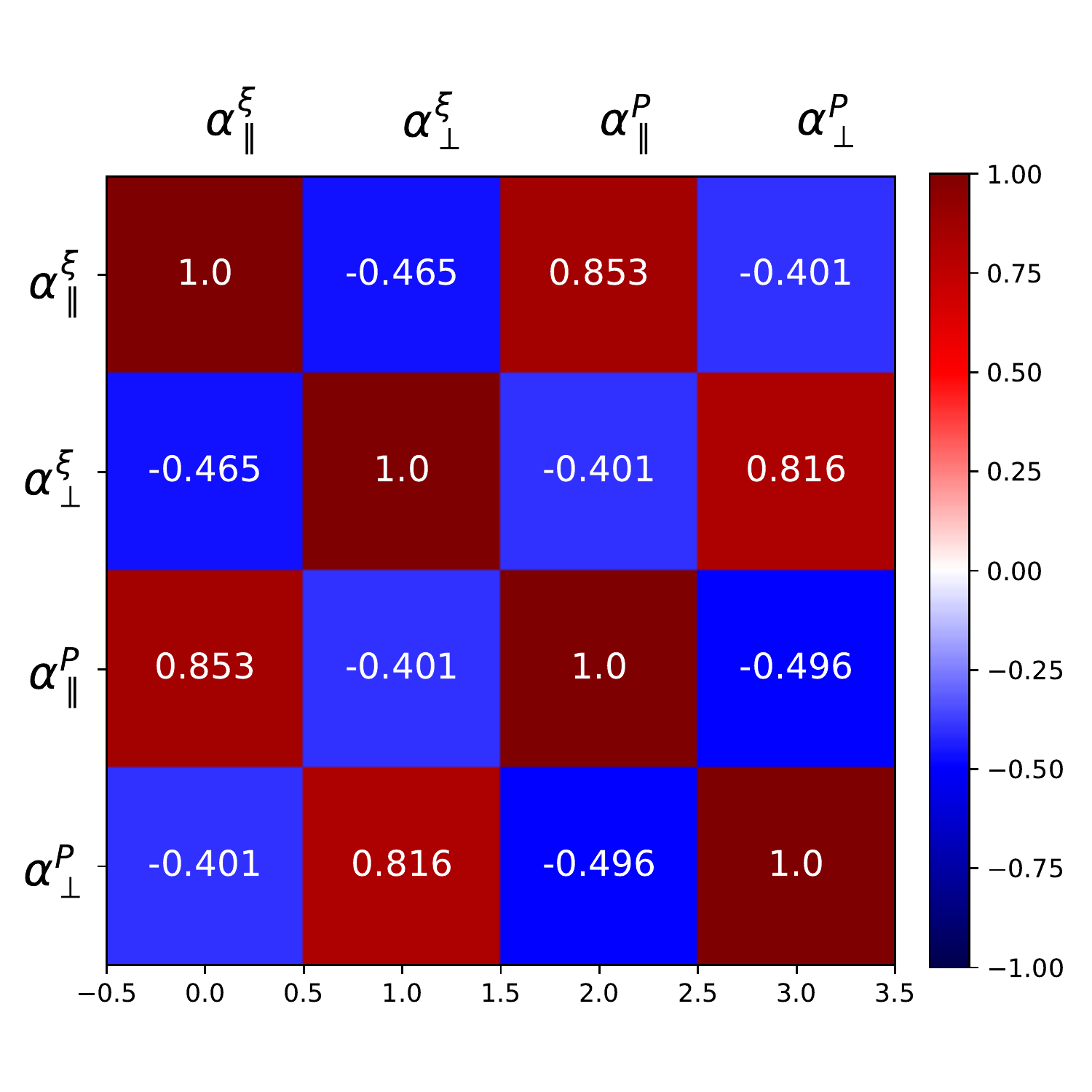}
	\caption{Correlation coefficients between $\alpha_\parallel$ and $\alpha_\perp$ in CS and FS obtained from fits to the prereconstructed 1000 EZmock realisation.}
	\label{fig:alpha_corr_tot_pre}
\end{figure}

Figure \ref{fig:covariance_pre} shows the correlation matrix between the prereconstructed 2PCF and PS. The main differences between prereconstruction and postreconstruction covariance matrices are in the off-diagonal blocks within each space (either FS or CS), which are significantly reduced in postreconstruction. We construct the $\alpha$'s parameters prereconstruction total covariance matrix obtained from the 1000 best-fit $(\aperp, \apara)$. The corresponding correlation matrix before the individual adjustments is shown in Figure \ref{fig:alpha_corr_tot_pre}. The reconstruction procedure enhances the correlations for a given parameter in two spaces, but reduces the correlations between the two parameters themselves (both in a given space and across spaces). Overall, the measurements depend on the signal-to-noise ratio of the BAO features, which are lower for prereconstruction data.

Table~\ref{tab:stats_prerecon} summarises the statistical properties of 
$\left(\alpha_\parallel, \alpha_\perp \right)$ for the different analyses,
CS, FS, GA and JS, performed on 1000 EZmock realisations. 
For each parameter, we show six quantities: the average bias $\Delta_\alpha = \langle \alpha - \alpha_{\rm exp} \rangle$ with respect to the
expected value (${\aperp}_{\rm , exp}={\apara}_{\rm , exp}=1$), 
the standard deviation of best-fit values $\sigma$, 
the mean estimated error $\langle \sigma_i\rangle$, 
the mean asymmetry of the estimated error distribution 
$\langle A\left(\sigma_i\right)\rangle =\langle 2\left(\sigma_{\rm i}^{\rm sup}-\sigma_{\rm i}^{\rm inf}\right)/\left(\sigma_{\rm i}^{\rm sup}+\sigma_{\rm i}^{\rm inf}\right)\rangle $,
the mean of the pull $Z_i = \left(\alpha_i - \langle \alpha_i\rangle \right)/\sigma_i$ 
and its standard deviation. 
If errors are correctly estimated and follow a Gaussian distribution, 
we expect that $\sigma = \langle\sigma_i\rangle$, $\langle Z\rangle=0$ and 
$\sigma(Z_i)=1$. For both parameters, the JS analysis presents larger systematic shifts $\Delta_\alpha$ and larger errors than the GA. For the four analysis, $\sigma(Z_i)$ is larger than one, indicating an underestimation of the errors. This effect is more important for the GA. As our modelling is not well suited to describe the non linearities of the prereconstruction two point statistics, for every analysis the precision and accuracy are degraded. Here the JS method takes into account the correlations between two observables that are not well modeled in the first place. However, using an appropriate full shape modeling and removing the broadbands nuisance parameters, the CS and FS should be described by the exact same set of parameters. Hence the JS could improve the analysis.

\begin{table*}
    \centering
    \caption{Statistics on the fit of the 1000 EZmocks prereconstruction realisations. $N_{\rm good}$ is the number of valid realisations after removing undefined contours and extreme values and errors. We show the mean value of the best-fit reduced ${}_r\chi^2_{\rm min}$. For each parameter, we show the average bias $\Delta_\alpha \equiv \langle \alpha_i - \alpha_{\rm exp}\rangle$, the standard deviation of best-fit values $\sigma \equiv \sqrt{\langle \alpha_i^2 \rangle - \langle \alpha_i\rangle^2}$, the average of the per-mock estimated uncertainties $\langle \sigma_i \rangle$, the asymmetry of the estimated error distribution $ A\left(\sigma_i\right) \equiv \langle 2\left(\sigma_{\rm i}^{\rm sup}-\sigma_{\rm i}^{\rm inf}\right)/\left(\sigma_{\rm i}^{\rm sup}+\sigma_{\rm i}^{\rm inf}\right)\rangle $, where $\sigma_i^{sup}$ and $\sigma_i^{\rm inf}$ are the superior and inferior one sigma errors estimated from the likelihood profile, the average of the pull $Z_i \equiv (\alpha_i - \langle \alpha_i \rangle)/\sigma_i$ and its standard deviation $\sigma(Z_i)$.}
    {
    \begin{tabular}{lcc|cccccc|cccccc}
    \hline
    \hline
    & & & \multicolumn{6}{c|}{$\alpha_{\perp}$}  & \multicolumn{6}{c}{$\alpha_{\parallel}$} \\
     & $N_{\rm good}$ & $\langle{}_{r}\chi^2_{\rm min}\rangle$ &
    $\Delta_\alpha $ &
    $\sigma $ & 
    $\langle\sigma_i\rangle$ & 
    $\langle A\left(\sigma_i\right)\rangle$ & 
    $\langle Z_i\rangle$ & 
    $\sigma(Z_i)$ & 
    $\Delta_\alpha $ &
    $\sigma $ & 
    $\langle\sigma_i\rangle $ & 
    $\langle A\left(\sigma_i\right)\rangle$ &
    $\langle Z_i \rangle$ & 
    $\sigma(Z_i)$ \\ 
    & & & [$10^{-2}$] & [$10^{-2}$]  & [$10^{-2}$] & [$10^{-2}$] & & 
        & [$10^{-2}$] & [$10^{-2}$] & [$10^{-2}$] & [$10^{-2}$] & & \\
    \hline

    CS& 891 & 0.961& -0.683& 3.243& 3.075& 5.575& -0.053& 1.048& -0.663& 5.794& 5.057& 4.237& -0.015& 1.107 \\
    FS& 945 & 1.005& -0.434& 3.261& 2.822& 3.79& -0.05& 1.104& 0.114& 6.153& 4.78& 0.646& 0.011& 1.187 \\
    GA& 876& -& -0.37& 2.954& 2.67& -& -0.044& 1.121& -0.196& 5.259& 4.366& -& -0.007& 1.208 \\ 
    JS& 931 & 0.959& -0.478& 3.373& 2.899& 5.061& -0.047& 1.084& 0.58& 6.533& 4.87& 0.21& 0.008& 1.19 \\

    \hline 
    \hline
    \end{tabular}
    }
    \label{tab:stats_prerecon}
\end{table*}

\section{Modelisation of the cross correlation pattern}
\label{sec:annex:cross_cov}

We propose an analytical description of the cross correlations between the correlation function and power spectrum multipoles. The 2-point statistic between the fluctuation field in configuration and fourier space is
\begin{equation}
\begin{split}
  \langle \delta_g(\vec{k})\delta_g(\vec{r}) \rangle &=\frac{1}{(2\pi)^3}\int d^3k'  e^{i \vec{k'}\cdot \vec{r}} \langle \delta_g(\vec{k})\delta_g(\vec{k'}) \rangle \\
  &=e^{i \vec{k}\cdot \vec{r}} P(\vec{k}),
\end{split}
\label{eq:cross_2pt}
\end{equation}
while assuming gaussian fields, the cross covariance is given by
\begin{equation}
\begin{split}
  Cov\left[P(\vec{k}),\xi(\vec{r})\right] &= \langle \delta_g(\vec{k})\delta_g(-\vec{k})\delta_g(\vec{r})\delta_g(-\vec{r}) \rangle
  \\&=\langle \delta_g(\vec{k})\delta_g(\vec{r})\rangle\langle \delta_g(-\vec{k})\delta_g(-\vec{r}) \rangle +\\& \quad \langle \delta_g(\vec{k})\delta_g(-\vec{r})\rangle\langle \delta_g(-\vec{k})\delta_g(\vec{r}) \rangle \\&= 2\cos (2\vec{k} \cdot \vec{r})|P(\vec{k})|^2.
\end{split}
\label{eq:cross_cov}
\end{equation}

Then, including the shot noise contribution $P_N$ and the volume rescaling, the cross covariance between the multipoles $P_l(k)$ and $\xi_{l'}(r)$ can be written
\begin{equation}
\begin{split}
 Cov & \left[P_l(k),\xi_{l'}(r)\right] = \frac{2l+1}{2}\frac{2l'+1}{2}\frac{1}{Vs}\int_{-1}^{1}d\mu_k L_l(\mu_k) |P(k,\mu_k)+P_N|^2 \\
 &\times \int_{-1}^{1}d\mu_r L_l(\mu_r) 2\cos\left(2rk\mu_k\mu_r+2kr\sqrt{1-\mu_k^2}\sqrt{1-\mu_r^2}\right).
\end{split}
\label{eq:cross_cov}
\end{equation}

\begin{figure}[b]
	\includegraphics[width=.8\columnwidth]{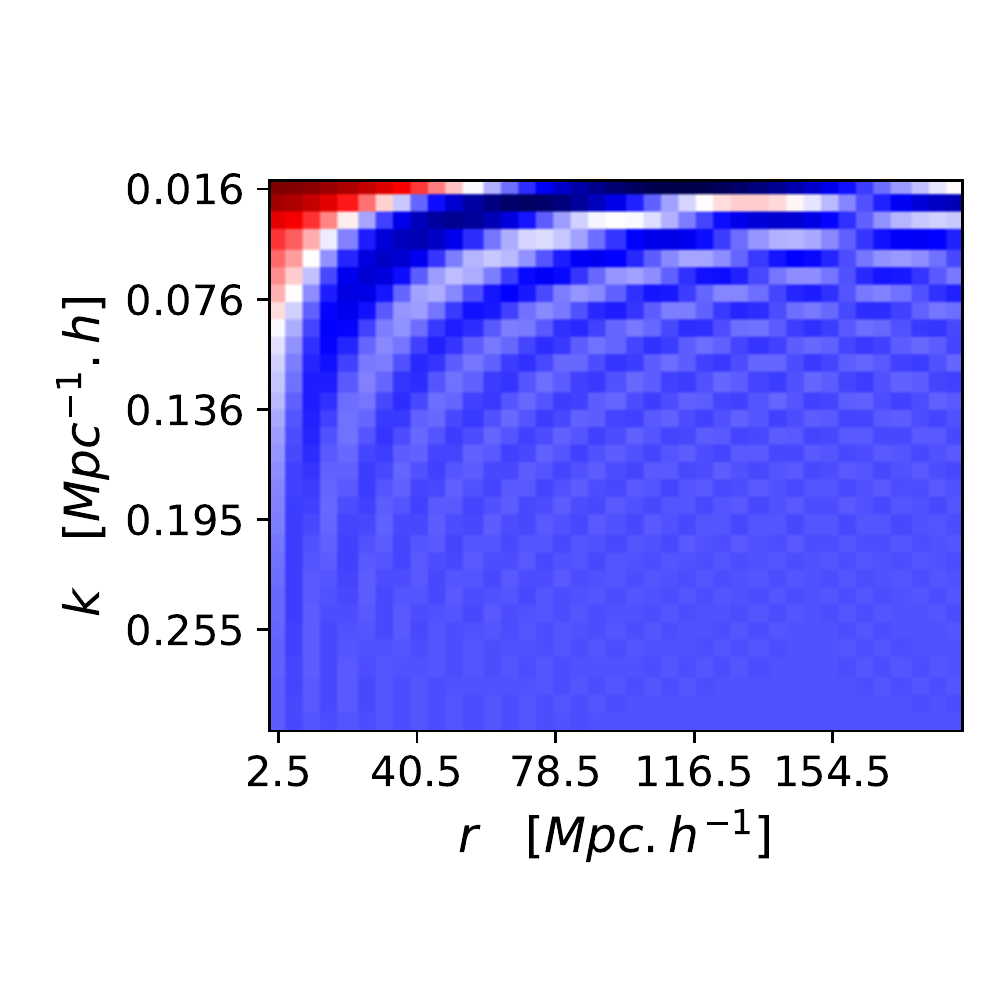}
	\caption{Analytical modeling of the cross correlations between the correlation function and the power spectrum monopoles for a given range of scales. We do not display the colorbar as we set the shot noise to zero and the volume of the survey to one. }
	\label{fig:cross_cov}
\end{figure}

Figure \ref{fig:cross_cov} shows the modeling of the cross covariance features using Eq \ref{eq:cross_cov} for a given range of scales. We find this description satisfying as we only aim for a physical understanding of these patterns, and use the full mocks sample to estimate the covariance. A more reliable modeling would require bin averaged integral as performed in \cite{grieb_gaussian_2016}.

\section{Mock dependant constraints}
\label{app:nmock}

We tested the robustness of the different analysis against the number of available mock realisations. We fit the averaged stack of mocks for different values of $N_{\rm mocks}$ used to estimate the sample covariance matrix. 

\begin{figure}[!h]
\begin{subfigure}[t]{0.9\columnwidth}
  \centering
  \includegraphics[width=\columnwidth]{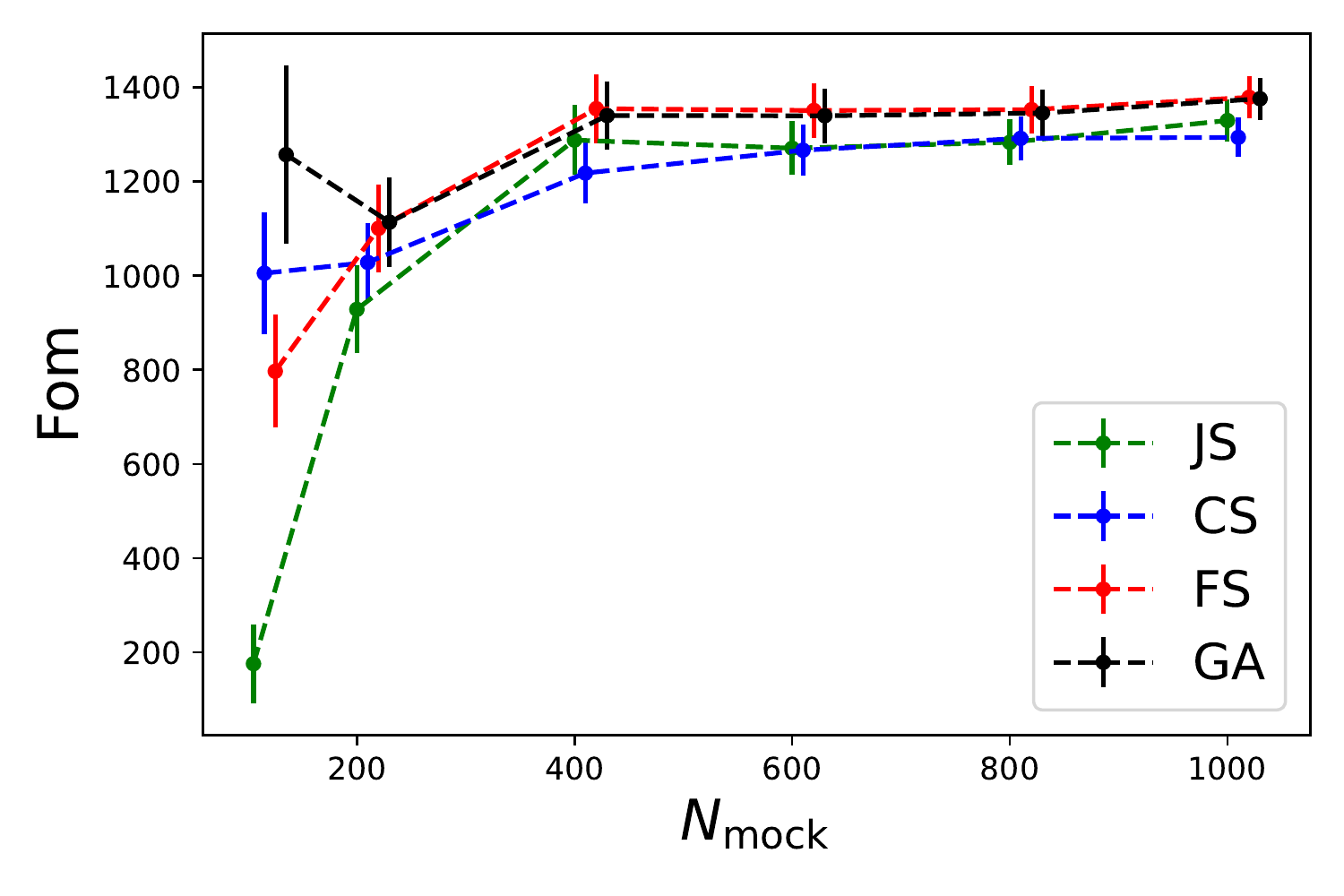}  
\end{subfigure}
\begin{subfigure}[t]{0.9\columnwidth}
  \centering
  \includegraphics[width=\columnwidth]{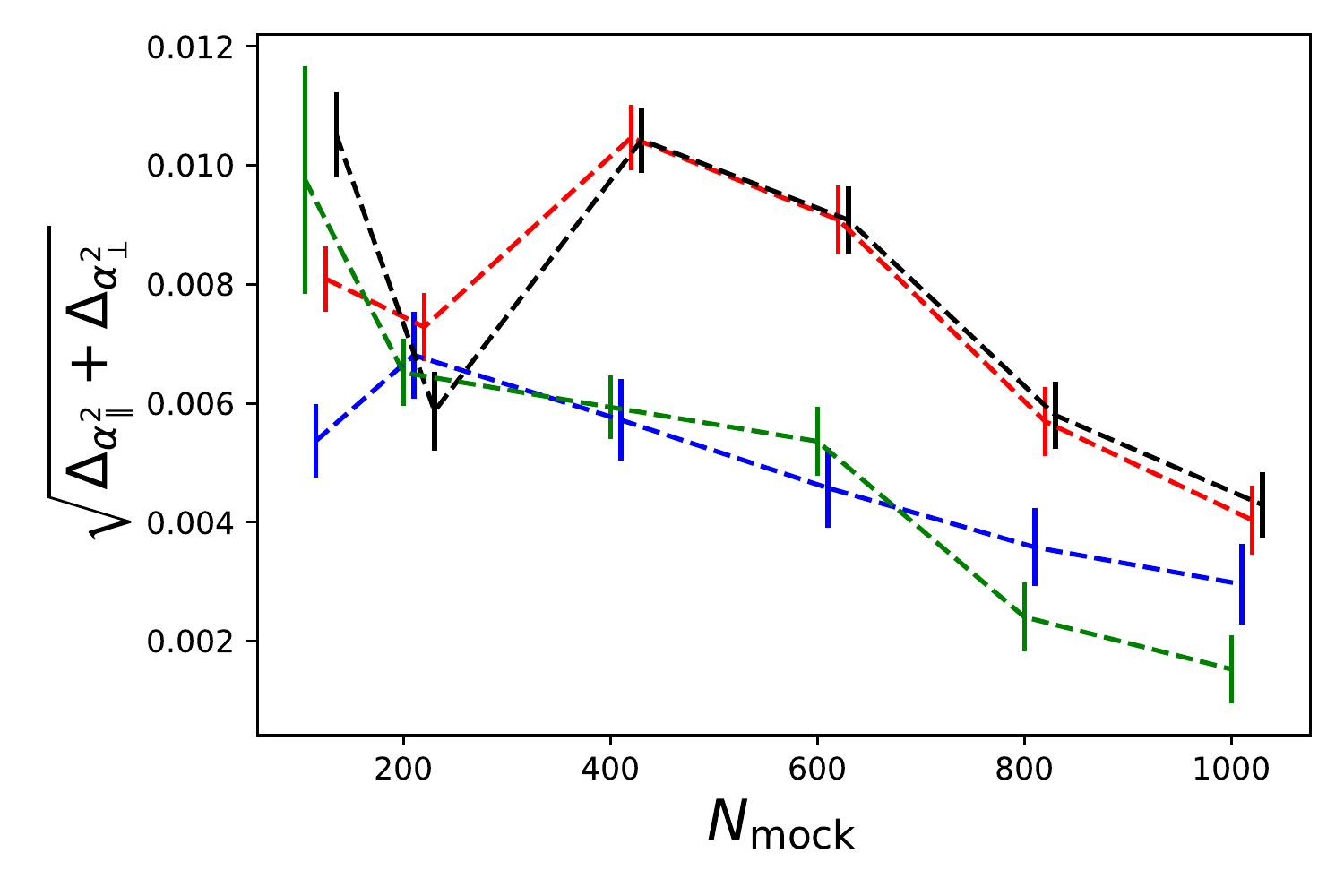}  
\end{subfigure}
\caption{Figure of merit (Fom) and quadratic shift from the expected parameters resulting from the fit of the stack of EZmocks for different $N_{\rm mock}$ used to estimated the sample covariance. Errors for the $\alpha$'s are scaled with $\sqrt{N_{\rm mock}}$ and errors for the Fom are derived from Eq 68 of \cite{taylorPuttingPrecisionPrecision2013}.}
\label{fig:precision_accuracy}

\end{figure}

Figure \ref{fig:precision_accuracy} shows the figure of merit defined as $Fom=1/\sqrt{|C_\theta|}$ with $C_\theta$ the covariance matrix of the inferred parameters, and the quadratic shift from the expected parameters $\sqrt{\Delta_\apara^2 + \Delta_\aperp^2}$ for different $N_{\rm mock}$. As expected we find that, due to the Whishart bound, the JS precision highly decreases when the number of mocks get small, typically the Fom drops when $N_{\rm mock}<4N_{\rm bins}$. However the JS constraint is compatible with the GA at 1$\sigma$, and appears to be robust as the Fom does not decrease very much in the range 1000-600 $N_{\rm mock}$. While the GA analysis introduces an additional dependence on $N_{\rm mock}$ through the covariance $C_c$ defined in Eq \ref{eq:C_c} (which has to be calculated for each point using the set of best fit parameters found with the appropriated sample covariance), the Fom is only decreased by a few percent when using 100 realisations. 

We find that for the JS the quadratic shift from the expected parameters monotonically decreases as the amount of mocks increases. For a large enough amount of realisations, the JS gives the smaller systematic bias.

\end{document}